\documentclass[journal]{IEEEtran}
\usepackage{booktabs}
\usepackage{cite}
\usepackage[pdftex]{graphicx}
\usepackage[cmex10]{amsmath}
\usepackage{amsmath,amsfonts,amssymb}
\usepackage{esint}
\usepackage{lineno}
\usepackage{subfigure}
\usepackage{algorithmic}
\usepackage[ruled,lined,boxed]{algorithm2e}
\usepackage{xcolor}
\usepackage{multirow}
\usepackage{diagbox}
\usepackage{threeparttable}
\usepackage{setspace}
\usepackage[justification=centering]{caption}
\usepackage[ruled,lined,boxed]{algorithm2e}
\usepackage{makecell}

\usepackage{hyperref}
\usepackage{geometry}

\newtheorem{remark}{Remark}

\begin{document}

\begin{spacing}{1}
	
\title{Integrated Location Sensing and Communication for Ultra-Massive MIMO With Hybrid-Field Beam-Squint Effect}
\author{
	Zhen Gao,~\IEEEmembership{Member,~IEEE}, Xingyu Zhou, 
	Boyu Ning,~\IEEEmembership{Member,~IEEE},
	Yu Su,~\IEEEmembership{Member,~IEEE},
	Tong Qin,\\
	Dusit Niyato,~\IEEEmembership{Fellow,~IEEE}
} 
	

\maketitle
\begin{abstract}
	
The advent of ultra-massive multiple-input-multiple-output systems holds great promise for next-generation communications, yet their channels exhibit hybrid far- and near- field beam-squint (HFBS) effect.
{In this paper, we not only overcome but also harness the HFBS effect to propose an integrated location sensing and communication (ILSC) framework.}
During the uplink training stage, user terminals (UTs) transmit reference signals for simultaneous channel estimation and location sensing. 
This stage leverages an elaborately designed hybrid-field projection matrix to overcome the HFBS effect and estimate the channel in compressive manner. 
Subsequently, the scatterers' locations can be sensed from the spherical wavefront based on the channel estimation results.	
By treating the sensed scatterers as virtual anchors, we employ a weighted least-squares approach to derive UT's location. 
Moreover, we propose an iterative refinement mechanism, which utilizes the accurately estimated time difference of arrival of multipath components to enhance location sensing precision. 
In the following downlink data transmission stage, we leverage the acquired location information to further optimize the hybrid beamformer, which combines the 
beam broadening and focusing to mitigate the spectral efficiency degradation resulted from the HFBS effect.
Extensive simulation experiments demonstrate that the proposed ILSC scheme has superior location sensing and communication performance than conventional methods. 

\end{abstract}


\IEEEpeerreviewmaketitle

\section{Introduction}

\IEEEPARstart{T}{he} next-generation wireless communication is poised to 
enhance the traditional mobile-broadband and real-time experiences to the next level and unlock novel use cases such as spatial perception and immersive platform  \cite{qc}.
To support these services, it is expected that the new sixth-generation (6G) targets will push existing key performance indicators further in areas such as throughput, latency, and coverage.
In this context, revolutionary physical layer (PHY) technologies serve as fundamental building blocks for the performance leap, among which the advanced multiple-input
multiple-output (MIMO) technology \cite{mmwave,YangWang, zjy} and spectrum expansion \cite{mmwave,wzw} hold great promise.

On the one hand, the evolution toward ultra-massive MIMO (UM-MIMO) emerges as a viable solution for further enhancing spatial degrees of freedom, thereby providing higher spectral efficiency. 
On the other hand, underutilized spectrum bands, such as millimeter-wave (mmWave) and terahertz (THz), harbor a substantial reservoir of available bandwidth, which enables the establishment of extremely wideband channels with unprecedented transmission {rates}. 
Moreover, the shorter wavelengths at these higher frequency bands permit the dense packing of more antennas within the same aperture size, effectively compensating for most of the path loss attenuation by resorting to highly directional beams \cite{wzw}. 
Consequently, UM-MIMO and mmWave (THz) technologies mutually complement each other, constituting a foundation of the 6G PHY.
   
Despite the appealing advantages of mmWave UM-MIMO systems, their channels exhibit the Hybrid Far- and Near- Field Beam-Squint (HFBS) effect due to the substantial increase in the number of antennas and the corresponding bandwidth, urgently necessitating a significant shift in signal processing paradigms.
Specifically, electromagnetic wave radiation can be categorized into two regions: Fraunhofer (far-field) and Fresnel (near-field), with their boundary quantified by the Rayleigh distance \cite{Fresnel}. 
The limited antenna aperture of current cellular networks makes the Rayleigh distance negligible for typical cellular networks. 
Consequently, existing communication technologies predominantly rely on the far-field radiation theories \cite{wzw-ISAC}.
However, as the antenna aperture sizes dramatically increase in the UM-MIMO systems, the near-field region expands rapidly, and some of the served user terminals (UTs) would fall into the near-field region and some otherwise, leading to the coexistence of the near-field and the far-field propagation \cite{hybrid-field,lzr}. 
Moreover, the combination of large bandwidth and UM-MIMO introduces a non-negligible propagation delay gap at different antennas for the same received signal filling this array aperture, which could be as large as multiple symbol periods.
This phenomenon is termed as the delay-squint effect \cite{lzr,gff2,law}. 
From the frequency domain perspective, the delay-squint effect could further introduce the beam-squint effect. 
To be more specific, the beam direction is a function of the operating frequency, and the undesired beam misalignment for the signals at marginal carrier frequencies occurs that reduces the spatial-frequency consistency \cite{gff2}.

Furthermore, with their broader bandwidth and larger antenna aperture, mmWave UM-MIMO systems inherently yield improved delay and angle resolution. 
Meanwhile, the hybrid-field propagation effect offers the potential of simultaneous angular and range resolutions by measuring the different antenna elements' phase difference generated by the 
spherical electromagnetic wave.
As a result, the HFBS effect of mmWave UM-MIMO systems also can be promising solutions for applications such as target localization \cite{lzr}, radar sensing \cite{near-field radar}, simultaneous localization and mapping (SLAM) \cite{SLAM}, and ultimately, integrated sensing and communication (ISAC) \cite{near-field ISAC}.

\subsection{Prior Work}

\begin{table*}

\renewcommand{\arraystretch}{1.5}
\centering
\begin{tabular}{|c|c|c|c|c|c|}
\hline
\multirow{2}{*}{Reference} & \multirow{2}{*}{Channel} & \multirow{2}{*}{Beam Squint} & \multirow{2}{*}{Configuration} & \multicolumn{2}{c|}{Functionality} \\
\cline{5-6}
 &  &  &  & Localization & Communication \\
 \hline
\cite{near-field radar} & Near field &  & Single base station & $\checkmark$ & \\
\hline
\cite{SLAM} & Near field &  & Single base station & $\checkmark$ & \\
\hline
\cite{cmy beam-squint} & Near field & $\checkmark$ & Single base station &  & $\checkmark$ \\
\hline
\cite{beam-squint CE} & Near field & $\checkmark$ & Single base station+RIS &  & $\checkmark$ \\
\hline
\cite{near-field music} & Near field &  & Single base station & $\checkmark$ & \\
\hline
\cite{near-field music2} & Near field &  & Single base station & $\checkmark$ & \\
\hline
\cite{near-field tracking} & Near field &  & Single base station & $\checkmark$ & \\
\hline
\cite{Near-Field Localization} & Near field &  & Single base station+RIS & $\checkmark$ & \\
\hline
\cite{lzr} & \makecell{Hybrid far\\and near field} & $\checkmark$ & Single base station+RIS & $\checkmark$ & $\checkmark$ \\
\hline
Our work & \makecell{Hybrid far\\and near field} & $\checkmark$ & Single base station & $\checkmark$ & $\checkmark$ \\
\hline
\end{tabular}
\captionsetup{labelfont={color=black}, textfont={color=black}}
\caption {Summary of existing works on ILSC}
\label{tab:ilsc_summary}
\end{table*}

Hybrid beamforming architecture has been widely considered for massive MIMO and UM-MIMO systems due to their low hardware cost achieved by using a much smaller number of radio frequency (RF) chains than antenna elements. \cite{omp-ce}. 
Nevertheless, this architecture significantly decreases the effective baseband measurement dimension of received signals, bringing great challenges to full-dimensional channel state information (CSI) acquisition.
To this end, compressive sensing (CS) technique, as an effective tool to recover the sparse signals via underdetermined measurements, has attracted extensive attention
for channel estimation in mmWave and THz bands \cite{omp-ce,kml,ShicongLiu}. 
Under the assumption of the far-field planar wavefront, the mmWave massive MIMO channel exhibits an obvious sparsity pattern in the angular domain.
Therefore, the channel estimation can be formulated as a larger number of antennasal sparse signal recovery problem, and solved resorting to the off-the-shelf CS-based algorithms \cite{omp-ce,init, MalongKe_IoTJ}.
However, when extended to UM-MIMO systems, the far-field approximation {no longer holds}, leading to a non-negligible energy dispersion and deteriorating its inherent sparsity.
For this purpose, the authors of \cite{myk} introduced a subarray piecewise approximate far-field-based sparse channel estimation scheme, which handled the channels of different subarrays independently with the far-field approximation.
However, it overlooked the geometrical correlations of different subarrays, thus increasing the parameters to be estimated, and only performing well at a moderate medium antenna array size.  
Furthermore, a more complex Fresnel approximation was developed based on second-order Taylor expansion, and \cite{cmy,near-field ce} proposed higher-dimensional projection matrices to enable sparse channel estimation at the cost of increased computational complexity. This is due to the reason that candidate columns steering
vectors of projection matrices were determined by both the angular and the range parameters.

Most of the aforementioned schemes have focused on leveraging the structured common sparsity of the spatial-frequency channel to enhance estimation performance, yet neglecting the beam-squint effect. 
To address this issue, \cite{cmy beam-squint} revealed the bilinear pattern of HFBS, which implies that the sparse support set of channels in both the angle and the range domains can be regarded as a linear function against frequency, and
developed a bilinear pattern based structured CS algorithm for channel estimation.
Additionally, \cite{lzr,beam-squint CE, KuiyuWang} developed a frequency-dependent projection matrix to correct the beams-squint effect and maintained the spatial-frequency consistency in the corresponding transform domain.
Besides, for the problem of dispersed beams caused by the beam-squint effect,  
a common solution is to employ true-time-delay (TTD) to replace phase shifters to generate frequency-dependent phase shift but inducing higher hardware cost \cite{law}. 
To retain the original cost-effective hybrid beamforming architecture, \cite{gff} proposed analog beam broadening strategies to divide the whole array into several virtual subarrays to generate a wider beam and provide constant beamforming gain as much as possible over the whole operating frequency band, thus mitigating the impact of far-field beam-squint to some extent. 
Furthermore, \cite{misfocus2} studied a spatial frequency modulated continuous wave (FMCW)-based phase shift design for massive phased arrays to overcome the near-field beam misfocus problem, and achieve a uniform gain over a wide bandwidth and a higher rate than the standard design in the near-field propagation region.

{
In 6G mobile communication systems, the use of higher frequency bands, wider bandwidth, and larger number of antennas will pave the way for advanced sensing capabilities with high accuracy and resolution.
As a result, future 6G ISAC systems are likely to support a variety of sensing applications, including ultra-high accuracy localization and tracking, simultaneous imaging, mapping, and localization, as well as enhanced human sensing, gesture, and activity recognition\cite{Bayesteh2022}.
Localization is a typical application of ISAC.
Therefore, the concept of integrated localization and sensing communication (ILSC) is proposed to further specify and refine the broader concept of ISAC.} 
In the context of sensing and localization, research efforts towards SLAM and ISAC techniques are well underway.
Standardized cellular network localization methods allocate dedicated reference signals, namely the downlink positioning reference signal (PRS) and uplink sounding reference signal (SRS), to facilitate positioning \cite{5G positioning}. 
These signals are instrumental in deriving location information by measuring parameters such as reference signal received power (RSRP), time difference of arrival (TDoA), angle of departure (AoD), angle of arrival (AoA), and multi-cell round trip time (Multi-RTT). Moreover, these measurements further enable precise UT localization, achieving accuracy ranging from a few meters to a few decimeters depending on the specific deployment scenarios \cite{mmwave localization}. { In \cite{RIS_position_SIMO,RIS_position_MIMO}, the authors proposed a joint localization and location-aware beamforming scheme for a semi-passive RIS-assisted system for the single-user and multiple-user scenarios, where the angular parameters were utilized to position the UTs. However, only the LoS path was considered and the delay information was also dismissed.}
{ In \cite{qiaoli}, the author design a joint users' activity and channel estimation scheme under near-field scenario and sense the users' location with the TDoA and AoA of the LoS links between multiple subarrays and users.}
For the near-field propagation region of UM-MIMO systems, the accurate spherical wavefront presents a new way for sensing and localization.
Particularly, \cite{near-field music,near-field music2} pioneered to develope a modified two-dimensional version of the multiple signal classification (MUSIC) algorithm for localizing the signal sources with the spherical wavefront sampled by a passive sensor array. 
Then, \cite{near-field tracking} derived the theoretical error bound for target tracking by only harnessing the curvature of arrival in phase-difference measurements in the near-field region, and revealed that the range information tends to decrease with increasing target distance.
The authors in \cite{Near-Field Localization} further proposed a CS-based reconfigurable intelligent surfaces (RIS) aided localization scheme for near-field region depending on the double anchors.
Moreover, \cite{near-field radar} presented a near-field wideband sensing scheme, which unveiled that coherent processing of spatial domain and frequency measurements contributes to better sensing accuracy compared with  \cite{near-field music,near-field music2,near-field tracking,Near-Field Localization}.   

However, the aforementioned schemes necessitated the allocation of dedicated radio resources for positioning signals. 
To minimize resource overhead and achieve highly integrated functionality, \cite{SLAM} proposed to utilize the beam sweeping procedure in the broadcast synchronization signal block (SSB),
where the angular parameters can be extracted for reconstructing images of the propagation environment and providing reliable UT localization service simultaneously.
Yet, this method only peroformed well for indoor scenario with smooth concrete wall and required the accurate prior information of the initial positions of UTs and base station (BS), significantly limiting its application scope.
Additionally, \cite{lzr} introduced an integrated uplink channel estimation and localization scheme for the RIS-assisted UM-MIMO system, which synthetically exploited the TDoA and AoA information of hybrid-field spherical wave for accurate UT localization based on the estimated channels, and the localization results were further used to improve the channel estimation performance in turn.
However, this method relied on double anchors including the BS and RIS, and also overlooked the design of data transmission for achieving ISAC architecture.
Meanwhile, the authors  in \cite{near-field ISAC} optimized the ISAC signal to maximize the near-field sensing performance subject to a minimum communication rate while requires the priori target location information and  only considered the ideal line-of-sight (LoS) propagation condition.
{  In \cite{ISAC_Wu1}, the UAV-enabled ISAC systems are shown to effectively address the requirements of periodic sensing scenarios. In addition, the authors in \cite{ISAC_Wu1,ISAC_Wu2} have investigated the problem of resource allocation between communication and sensing in ISAC.}
{
Additionally, it may be beneficial to explore network-level cooperative ISAC, which leverages multi-cell cooperation to enhance both sensing and communication coverage. As discussed in \cite{2024_arXiv_Meng}, network-level ISAC can provide additional degrees of freedom for achieving better integration between sensing and communication, thereby improving overall system performance. By coordinating across multiple ISAC base stations, such a network can optimize resource allocation and interference management, offering promising opportunities for overcoming limitations in traditional ISAC setups. }

{
Based on the review above and Table I, there still exists a gap among previous works in investigating ILSC performance with only a  single UM-MIMO anchor and considering both HF and beam squint effects simultaneously.}  
 
%

 \subsection{Our Contributions}
 
In this work, we focus on achieving ILSC function for time-division duplexing (TDD) mmWave UM-MIMO systems. To be specific, the main contributions of this paper can be summarized as follows:    
\begin{itemize}
 	\item \textbf{We conceive  a novel frame structure and signal processing flow designed for ILSC. }
	The proposed architecture consists of a training stage and a data transmission stage. 
	During the training stage, UTs transmit the uplink reference signals to acquire CSI. 
	Moreover, without the need for additional radio resources, a BS can simultaneously sense the positions of scatterers based on the estimated channels and further provide accurate UT localization service by exploiting the hybrid-field propagation characteristic. 
	In the following downlink data transmission stage, the BS further employs location information to design beamforming schemes for more efficient payload data transmission.
 	 
	\item {\textbf{We utilize a projection-based Bayesian CS channel estimation scheme to combat the HFBS effect.} By utilizing an elaborate frequency-dependent hybrid-field projection matrix, we ensure that the channel vectors (from one UT's antenna to all BS's antennas) represented in the projected domain have the identical sparsity patterns across different sub-carriers and different antennas on the UT side. On this basis, we further consider the generalized multiple measurement vector approximate message passing (GMMV-AMP) algorithm to solve this CS problem, where the expectation-maximization (EM) technique is combined to adaptively learn the unknown hyper-parameters across all subcarriers.}
     
	\item  \textbf{We propose to simultaneously locate the scatterers and UT based on the acquired CSI, by leveraging the hybrid-field propagation characteristic of UM-MIMO systems.} 
 	Exploiting acquired CSI, we simultaneously determine locations of scatterers and a UT without requiring additional radio resources. 
 	Specifically, we extract the targets' angle and distance information from the spherical electromagnetic wavefront based on the estimated channels to sense the scatterers' locations. 	
 	Then, by treating these sensed scatterers as virtual anchors (VAs), we can coarsely estimate the UT's location according to the propagation geometry. 
 	To further enhance the sensing performance, we utilize super-resolution relative delay estimation so that the UT's and scatterers' location can be refined through an iterative procedure.
 
 	\item 
	\textbf{We introduce a hybrid beamforming design for downlink data transmission that leverages the location sensing results and overcomes the HFBS effect.}
	Specifically, we first develop an analog beamforming codebook.
	Within this codebook, each beamforming vector is associated with a pair of estimated distance and angle parameters, and thus can focus the signal energy on the sensed position under the hybrid-field effect. To further mitigate the beam-squint effect, each beam is designed to be broadened along both the distance and angle domains simultaneously to cover the beam-squint region, thus providing constant beamforming gain as much as possible over the whole bandwidth and enhance spectral efficiency (SE). Furthermore, we employ the simultaneous orthogonal matching pursuit (SOMP) algorithm to select optimal analog beamforming vectors and design the corresponding digital precoder to enable multi-stream transmission.
\end{itemize}

\subsection{Notations}

Throughout this paper, scalar variables are represented by normal-face letters, while column vectors and matrices are indicated by boldface lower and upper-case letters, respectively. 
The transpose, Hermitian transpose, and inversion of matrices are denoted by $(\cdot)^{\rm T}$, $(\cdot)^{\rm H}$, and $(\cdot)^{-1}$, respectively. 
In addition, $|\cdot|$ represents the modulus, $\Vert \cdot \Vert_p$ denotes the $\ell_p$-norm, and $\Vert \cdot \Vert_{\rm F}$ signifies the Frobenius-norm.
$[\cdot]_i$ denotes the $i$-th element of a vector, and $[\cdot]_{i,j}$ represents the $i$-th row and $j$-th column element of a matrix.
Furthermore, $|\cdot|_c$ indicates the cardinality of a set.
$\mathcal{CN}(\cdot,\mu,\Sigma)$ represents the Gaussian distribution with mean $\mu$ and variance $\Sigma$, while $\mathcal{U}[a,b]$ stands for a uniform distribution between $a$ and $b$. 
$\mathbb{E}(\cdot)$ and $\mathbb{V}(\cdot)$ calculate the mean and variance of a variable, respectively.
$\partial(\cdot)$ denotes the first-order partial derivative operation, and ${\rm diag}(\mathbf{a})$ constructs a diagonal matrix with elements from vector $\mathbf{a}$ placed along its diagonal.
Finally, $\mathbf{I}_n$ denotes the identity matrix of size $n \times n$, and $\mathbf{0}_{n\times m}$ represents the $n \times m$ zero matrix.

\section{System Model}

\begin{figure}[t]	
	\centering
	\includegraphics[width=0.8\columnwidth, keepaspectratio]{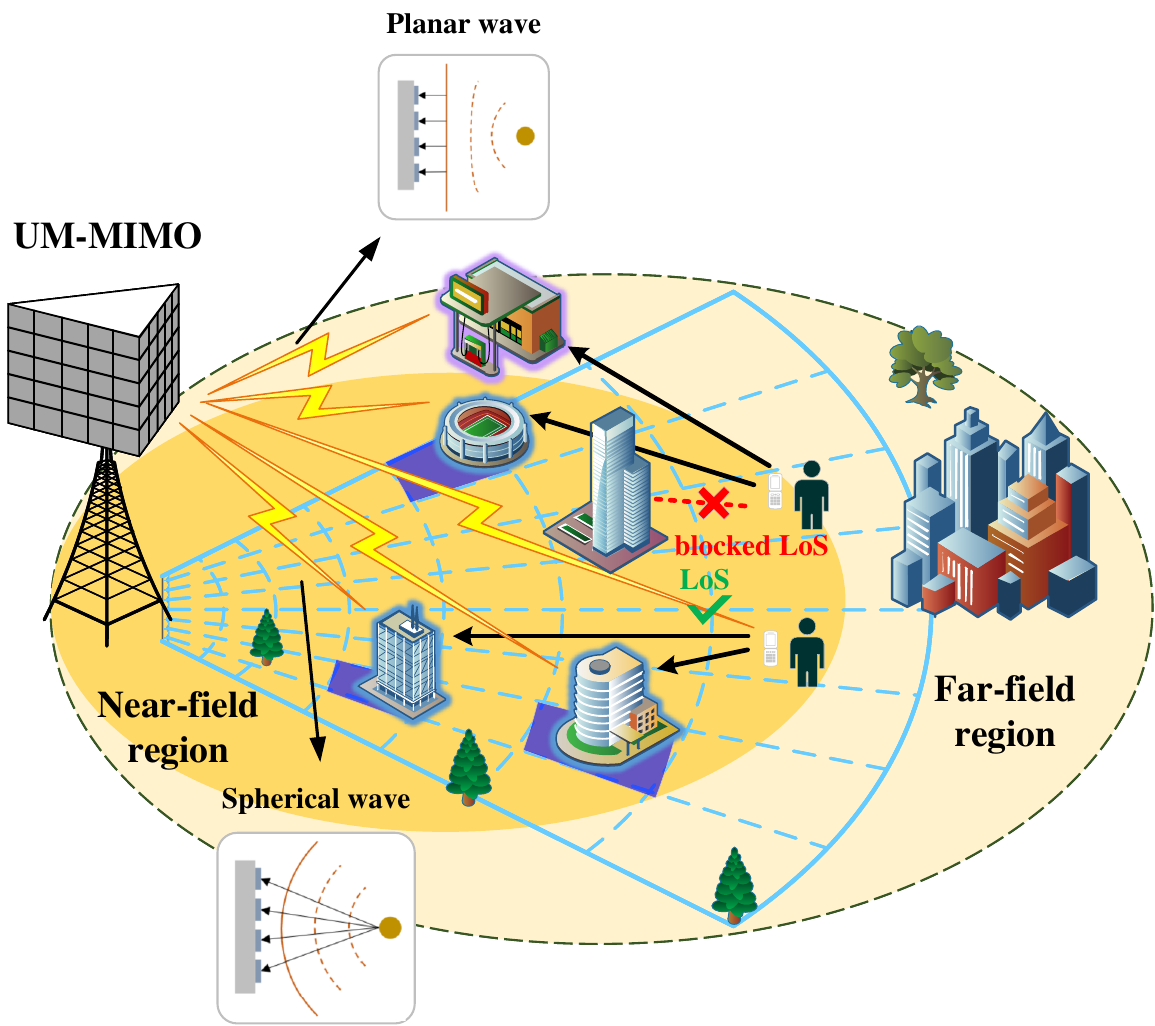}
	\captionsetup{font={color = {black}}, justification = raggedright,labelsep=period}
	\caption{The ILSC scenario for mmWave UM-MIMO systems.}
	\label{fig1:system model}	
	\vspace{-5mm}
\end{figure}

Consider an ILSC scenario in mmWave UM-MIMO systems as shown in Fig.~\ref{fig1:system model}, where a BS serves $U$ UTs in TDD mode.
The BS and UTs are all equipped with a uniform linear
array (ULA)\footnote{{If uniform planar
array is considered, the dimensionality of the parameters becomes too large to run the simulation.
Therefore, without changing the core of the problem, we have made a relatively idealized treatment of the practical problem in order to gain an understanding of the essential problem of HFBS effect in channel estimation and localization, and to gain corresponding insights.}}, and comprise $N_{\rm BS} \ (N_{\rm UT})$ antennas and $N_{\rm BS}^{\rm RF} \ (N_{\rm UT}^{\rm RF})$ RF chains.
To reduce the hardware costs and power consumption of UM-MIMO, the BS and UTs all employ hybrid beamforming architecture with a much smaller number of RF chains than that of antennas, i.e., $N_{\rm BS} \gg N_{\rm BS}^{\rm RF} \ (N_{\rm UT} \gg N_{\rm UT}^{\rm RF})$. 
Moreover, the cyclic prefix orthogonal frequency division multiplexing (CP-OFDM) with $M$ subcarriers is adopted to combat the frequency selective fading of the broadband mmWave channels.

\subsection{Uplink Reference Signal Transmission} \label{sec:2.1}

At this stage, multiple UTs transmit mutually orthogonal reference signals in the $Q$ consecutive time slots in the uplink to acquire CSI.  
For a specific UT, in order to estimate the $m$-th subcarrier’s CSI, the received baseband signal $\tilde{\mathbf{y}}_{\rm UL}[m,q] \in \mathbb{C}^{N_s^p}$ at the BS in the $q$-th time slot can be formulated as 
\begin{align}\label{eq:UL-CE}
	\tilde{\mathbf{y}}_{\rm UL}[m,q] = &  \frac{P_t^{\rm UL}}{M} \mathbf{W}^{\rm H}_{\rm UL}[m,q] \mathbf{H}_{\rm UL}[m] \mathbf{f}_{\rm UL}^{\rm RF}[q] s[m,q] \notag \\
	& +  \mathbf{W}^{\rm H}_{\rm UL}[m,q] \tilde{\mathbf{n}}[m,q],
\end{align}
where $1 \le q \le Q$ and $1 \le m \le M$. In \eqref{eq:UL-CE}, $\mathbf{W}_{\rm UL}[m,q]$ represents the BS's receive combining matrix $\mathbf{W}_{\rm UL}[m,q] = \mathbf{W}^{\rm RF}_{\rm UL}[q] \mathbf{W}^{\rm BB}_{\rm UL}[m,q] \in \mathbb{C}^{N_{\rm BS} \times N_s^p}$, in which $\mathbf{W}^{\rm RF}_{\rm UL}[m,q] \in \mathbb{C}^{N_{\rm BS} \times N_{\rm BS}^{\rm RF}}$ and $\mathbf{W}^{\rm BB}_{\rm UL}[m,q] \in \mathbb{C}^{N_{\rm BS}^{\rm RF} \times N_s^p}$ are the analog and digital receive combining matrices respectively, 
$N_s^p$ is the number of independent received signal streams on each subcarrier. 
By activating only one RF chain in one time slot for UT, the RF reference signal vector transmitted by the UT can be denoted as $\mathbf{f}_{\rm UL}^{\rm RF} [q] \in \mathbb{C}^{N_{\rm UT} \times 1}$.
$P_t^{\rm UL}$ is the total uplink transmit power, 
while $s[m,q]$ is the baseband reference signal with $s[m,q]s^*[m,q] = 1$.  
In addition, $\mathbf{H}_{\rm UL}[m] \in \mathbb{C}^{N_{\rm BS} \times N_{\rm UT}}$ stands for the corresponding uplink frequency-domain channel matrix, whose modeling approach will be detailed in Sec. \ref{sec:2.2}, and $\tilde{\mathbf{n}}[m,q] \in \mathbb{C}^{N_{\rm BS} \times 1}$ is the 
complex additive white Gaussian noise (AWGN) with the covariance matrix $\sigma^2 \mathbf{I}_{N_{\rm BS}}$, i.e., ${\mathbf{n}}[m,q] \sim \mathcal{CN}(\mathbf{0}_{N_{\rm BS}},\sigma^2 \mathbf{I}_{N_{\rm BS}})$.
Due to the constant modulus constraint of the adopted fully-connected RF phase shift network at the BS and UT, the uplink transmit and combining matrices can be expressed as 
$\left[\mathbf{f}_{\rm UL}^{\rm RF}[q]\right]_{i_1} = \frac{1}{\sqrt{N_{\rm UT}}} e^{j\vartheta^1_{i_1}}$ and  $\left[\mathbf{W}_{\rm UL}^{\rm RF}[q]\right]_{i_1,i_2} = \frac{1}{\sqrt{N_{\rm BS}}} e^{j\vartheta^2_{i_1,i_2}}$ respectively, where  $\vartheta^1_{i_1}$ and $\vartheta^2_{i_1,i_2}$ denote the phase value connecting the $i_1$-th antenna and the $i_2$-th RF chain, respectively.

\begin{figure*}[t]	
	\begin{align}\label{eq:channel model}
	 \mathbf{H}_{\rm UL}[m]  = \mathbf{H}_{\rm UL}^{\rm LoS}[m] + 
	\sum_{l=1}^{L} \sqrt{\frac{\beta_{\rm NLoS}^l N_{\rm BS} N_{\rm UT}}{L}} 
	 \alpha_l e^{-j\frac{2\pi}{\lambda_m}(r_l^{\rm UT}+r_l^{\rm BS})} \mathbf{a}_{\rm BS} (\theta_l^{\rm BS},r_l^{\rm BS},m) \mathbf{a}_{\rm UT}^{\rm H} (\theta_l^{\rm UT},r_l^{\rm UT},m),
	\end{align}
	\hrulefill
\end{figure*}

\subsection{Channel Model with HFBS Effect} \label{sec:2.2}

Considering the channel reciprocity in TDD systems, we focus on the formulation of the uplink channel matrix next.
In an mmWave UM-MIMO system with a hyper-scale antenna aperture and ultra-wide bandwidth, the channel exhibits an intrinsic HFBS effect.
{Our study specifically targets scenarios that are representative of indoor and short-range localization and communication environments, rather than urban macro-cell (UMa) environments. In these indoor settings, both near-field and far-field effects are present, along with significant beam squint effects due to the large antenna aperture and wide bandwidth, which are distinct from the purely far-field propagation considered in UMa scenarios.
}Under the spherical wavefront assumption, the spatial-frequency domain hybrid near- and far-field channel matrix $\mathbf{H}_{\rm UL}[m]$ can be accurately modeled as \eqref{eq:channel model} shown at the top of next page,
which is a sum of the contributions of $(L+1)$ multipath components (MPCs) including one LoS MPC and several non-LoS (NLoS) MPCs reflected by $L$ scattering clusters \cite{lyw}.
Particularly, for the LoS component $\mathbf{H}_{\rm UL}^{\rm LoS}[m]$, its $(n_1,n_2)$-th element can be further written as 
\begin{align}\label{eq:los}
	\left[ \mathbf{H}_{\rm UL}^{\rm LoS}[m] \right]_{n_1,n_2} = \sqrt{\beta_{\rm LoS} } e^{-j \frac{2\pi}{\lambda_m} r_{n_1,n_2}},
\end{align}
where $\beta_{\rm LoS}$ is the large-scale fading gain of the LoS link, $\lambda_m$ represents the wavelength of the $m$-th
subcarrier, and $r_{n_1,n_2}$ denotes the physical distance between the $n_1$-th antenna element of the BS and the $n_2$-th antenna element of the UT. 
\begin{figure*}[t]	
	\centering
	\includegraphics[width=1.9\columnwidth, keepaspectratio]{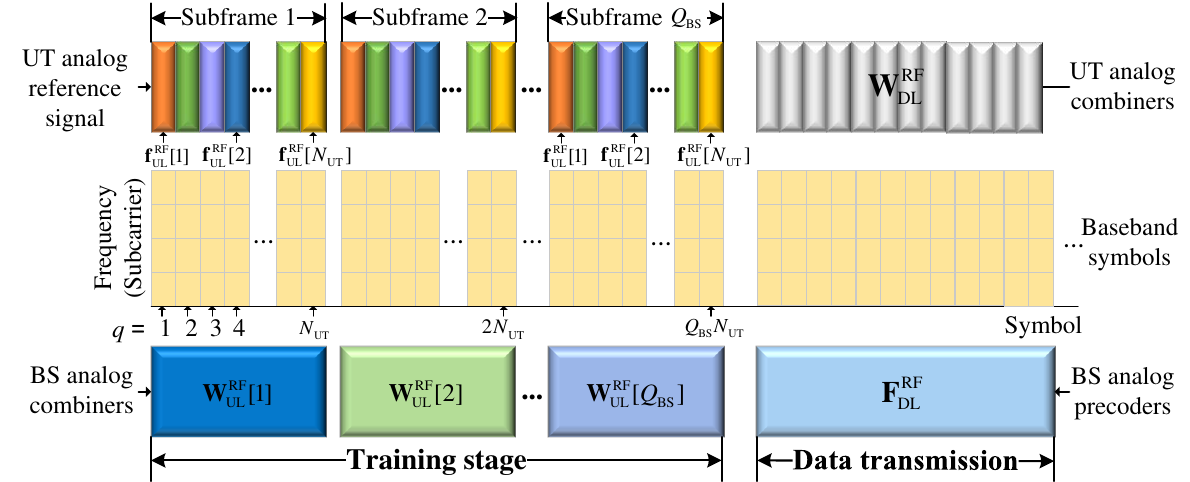}
	\captionsetup{font={color = {black}}, justification = raggedright,labelsep=period}
	\caption{The proposed frame structure for ILSC.}
	\label{fig2:frame structure}
	\vspace{-5mm}	
\end{figure*}
For the $l$-th NLoS component, it can be decomposed into the product of frequency-dependent array manifold of BS and UT sides under the spherical wave assumption, which can be further concretely presented as \cite{cmy}
\begin{align}\label{eq:BS array manifold}
	& \mathbf{a}_{\rm BS}  (\theta_l^{\rm BS},r_l^{\rm BS},m)  = \notag \\
	& \quad \frac{1}{\sqrt{N_{\rm BS}}}
	\left[e^{-j \frac{2\pi}{\lambda_m} (r_{l,1}^{\rm BS} - r_l^{\rm BS}) } , \dots ,  e^{-j \frac{2\pi}{\lambda_m} (r_{l,N_{\rm BS}}^{\rm BS} - r_l^{\rm BS}) }  \right]^{\rm T}.
\end{align} 
Herein, $r_l^{\rm BS}$ and $\theta_l^{\rm BS}$ denote the distance and AoA between the $l$-th scatterer and the reference point of BS's antenna array, respectively. Without loss of generality, the reference point could be set to the center of the antenna array.
More generally, $r_{l,n_1}^{\rm BS}$ represents the distance between the $l$-th scatterer and the $n_1$-th element of BS's antenna array. Under the spherical wave assumption, $r_{l,n_1}^{\rm BS}$ is a function of both $r_l^{\rm BS}$ and $\theta_l^{\rm BS}$, i.e.,       
$r_{l,n_1}^{\rm BS} = \sqrt{\left( {r_l^{\rm BS}} \right)^2 + \delta^2_{n_1} d^2 - 2 r_l^{\rm BS} \delta_{n_1} d \sin(\theta_l^{\rm BS}) }$, in which $\delta_{n_1} = (2n_1-N_{\rm BS}-1)/2$ with $n_1 = 1,2,\dots,N_{\rm BS}$, and $d$ denotes adjacent antenna spacing commonly with half-wavelength.
Similarly, $\mathbf{a}_{\rm UT} (\theta_l^{\rm UT},r_l^{\rm UT},m)$ can be expressed as 
\begin{align}\label{eq: UT array manifold}
	& \mathbf{a}_{\rm UT} (\theta_l^{\rm UT},r_l^{\rm UT},m) = \notag \\
	& \; \frac{1}{\sqrt{N_{\rm UT}}}
	\left[e^{-j \frac{2\pi}{\lambda_m} (r_{l,1}^{\rm UT} - r_l^{\rm UT}) } , \dots ,  e^{-j \frac{2\pi}{\lambda_m} (r_{l,N_{\rm UT}}^{\rm UT} - r_l^{\rm UT}) }  \right]^{\rm T},
\end{align} 
where $r_l^{\rm UT}$ and $\theta_l^{\rm UT}$ denote the distance and AoD between the reference point of UT's antenna array and the $l$-th scatterer, respectively.
Additionally, in \eqref{eq:channel model}, $\beta_{\rm NLoS}^l$ and $\alpha_l \sim \mathcal{CN}(0,\sigma_{\alpha}^2) $ are the large-scale fading gain of the NLoS link and small-scale channel gain corresponding to the $l$-th NLoS component, respectively. 

{Note that in practical environments, multi-hop paths experience significant path loss, as shown in \cite{NLoS-loss}, where each hop results in approximately 6 dB reflection loss and 12 dB transmission loss in a building scatterer scenario. Consequently, these multi-hop signals are treated as noise, having a minimal effect on both communication and sensing capabilities. Therefore, in the modeling presented in this paper, only single-hop NLoS components are considered.}

\section{Uplink Channel State Information Acquisition}

This section introduces the uplink CSI acquisition, i.e., the BS estimates the uplink channels based on the pilot signals transmitted by the UTs. 
Specifically, we first conceive a dedicated frame structure for ILSC and formulate the channel estimation problem as a structured CS problem.
Moreover, we conceive a frequency-dependent hybrid-field projection matrix to unveil the inherent sparse pattern of the channels with HFBS effect. 
Finally, we develop an algorithm that combines AMP with the EM technique to address this CS problem.

\subsection{Problem Formulation}

The designed frame structure for ILSC is depicted in Fig. \ref{fig2:frame structure}. Its time-frequency radio resources can be divided into multiple resource elements to convey both reference signals and payload data. Specifically, the first $Q = N_{\rm UT} Q_{\rm BS}$ symbols are allocated for channel training, while the remaining resources are reserved solely for payload data transmission.
During the full frame, the discrete Fourier transform (DFT) length of an OFDM symbol is set to $M_p = M_{\text{CP}}$. Consequently, the symbol duration of the reference signal becomes $(M_p + M_{\text{CP}})/B_w$, where $M_{\text{CP}}$ represents the CP length, and $B_w$ is the system bandwidth.
Furthermore, the reference signals in the training stage consists of $Q_{\text{BS}}$ subframes. Within each subframe, the BS analog receive combiner $\mathbf{W}_{\text{UL}}^{\text{RF}}[q]$ remains constant, and its phase shift of $(i_1,i_2)$-th elements $\vartheta^2_{i_1,i_2}$ for $1 \le i_1 \le N_{\rm BS}$ and $1 \le i_2 \le N_{\rm BS}^{\rm RF}$ are  uniformly distributed in $\mathcal{U} [0,2\pi)$.
Meanwhile, the digital receive combiner at the BS adopts an identity matrix $\mathbf{W}^{\text{BB}}_{\text{UL}}[m,q] = \mathbf{I}_{N_{\text{BS}}^{\text{RF}}}$, where $N_s^p = N_{\text{BS}}^{\text{RF}}$.
In this setting, UTs transmit $N_{\text{UT}}$ orthogonal reference signals in each subframe, which could be designed as a DFT matrix as $\mathbf{F}_{\text{UL}} = \left[ \mathbf{f}_{\text{UL}}^{\text{RF}}[1], \mathbf{f}_{\text{UL}}^{\text{RF}}[2],\dots,\mathbf{f}_{\text{UL}}^{\text{RF}}[N_{\text{UT}]} \right] \in \mathbb{C}^{N_{\text{UT}} \times N_{\text{UT}}}$ 
to meet the prerequisites of constant modulus and orthogonality.

Based on the above assumptions, by collecting $\tilde{\mathbf{y}}_{\rm UL}[m, (q_1-1)N_{\rm UT} + q_2]$ for all $Q_{\rm BS}$ subframes together (where $q_1$ is the subframe index and $q_2$ is the symbol index in one subframe), the aggregate received signals $\tilde{\mathbf{Y}}_{\rm UL}[m] \in \mathbb{C}^{P \times N_{\rm UT}}$ can be written as 
\begin{align}\label{eq:stacked signal}
	\tilde{\mathbf{Y}}_{\rm UL}[m] = \mathbf{W}^{\rm H}_{\rm UL} \mathbf{H}_{\rm UL}[m] \mathbf{F}_{\rm UL} \mathbf{S}[m] + \mathbf{W}^{\rm H}_{\rm UL} \tilde{\mathbf{N}}[m],
\end{align} 
where $\tilde{\mathbf{Y}}_{\rm UL}[m]$ successively selects every 
symbol of the same subframes $1 \le q_2 \le N_{\rm UT}$ as its columns, and     
stacks different subframes $1 \le q_1 \le Q_{\rm BS}$ on top of each another, and $P = Q_{\rm BS} N_{\rm BS}^{\rm RF}$ denotes the effective measurement length.  
Besides, $\mathbf{W}_{\rm UL} =  \left[  \mathbf{W}_{\rm UL}^{\rm RF} [1],  \mathbf{W}_{\rm UL}^{\rm RF} [2], \dots, \mathbf{W}_{\rm UL}^{\rm RF} [Q_{\rm BS}] \right] \in \mathbb{C}^{N_{\rm BS} \times P}$, $\mathbf{S}[m] = \sqrt{P_t^{\rm UL}/M} {\rm diag} \left\{ s[m,1],s[m,2],\dots,s[m,N_{\rm UT}] \right\} $ is the stacked baseband reference signal, and $\tilde{\mathbf{N}}[m] \in \mathbb{C}^{N_{\rm BS} \times N_{\rm UT}}$ stands for the corresponding AWGN for all these $Q_{\rm BS}$ subframes. Finally, the received baseband signal in \eqref{eq:stacked signal} is post-processed by
multiplying $\mathbf{S}^{\rm H}[m] \mathbf{F}_{\rm UL}^{\rm H}$ right-hand, i.e.,     
\begin{align} \label{eq:eff CE}
	{\mathbf{Y}}_{\rm UL}[m] = \sqrt \frac{M}{P_t^{\rm UL}} \tilde{\mathbf{Y}}_{\rm UL}[m] \mathbf{S}^{\rm H}[m] \mathbf{F}_{\rm UL}^{\rm H} = \mathbf{W}^{\rm H}_{\rm UL} \mathbf{H}_{\rm UL}[m]  + {\mathbf{N}}[m], 
\end{align} 
in which ${\mathbf{N}}[m] = \mathbf{W}^{\rm H}_{\rm UL} \tilde{\mathbf{N}}[m] \mathbf{S}^{\rm H}[m] \mathbf{F}_{\rm UL}^{\rm H}$ is the effective noise term. 
                                                    
\subsection{Projection Matrix Design} \label{sec:Projection Matrix Design}

\begin{figure*}[t]	
	\centering
	\subfigure[The representation of NLoS components with frequency-flat DFT projection matrix.]{
		\includegraphics[width=0.62\columnwidth, keepaspectratio]{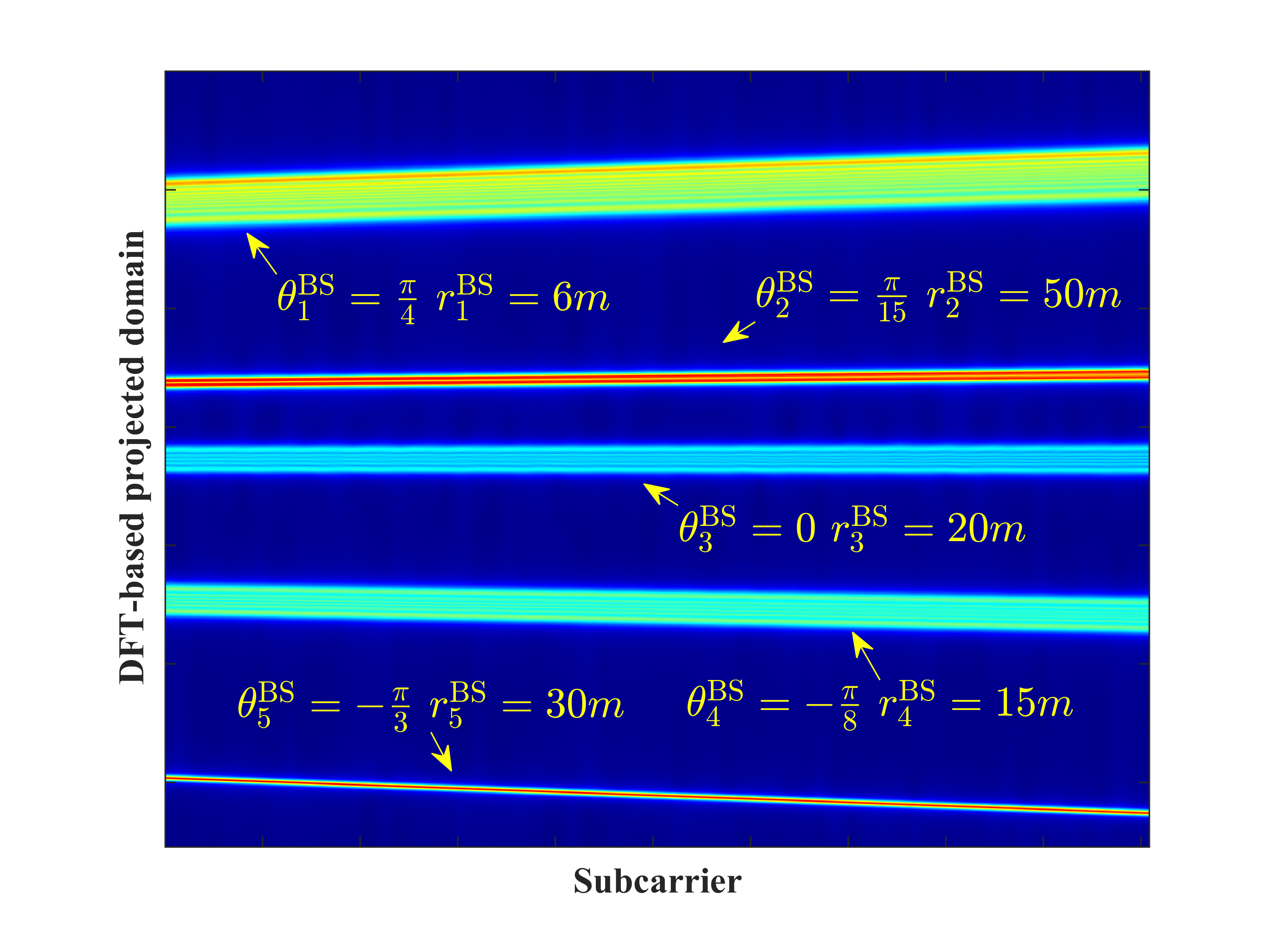}
		\label{fig3:dft}
	}
	\subfigure[{The representation of NLoS components with projection matrix defined in \eqref{eq:polar-domain projetion matrix}}.]{	
		\includegraphics[width=0.62\columnwidth, keepaspectratio]{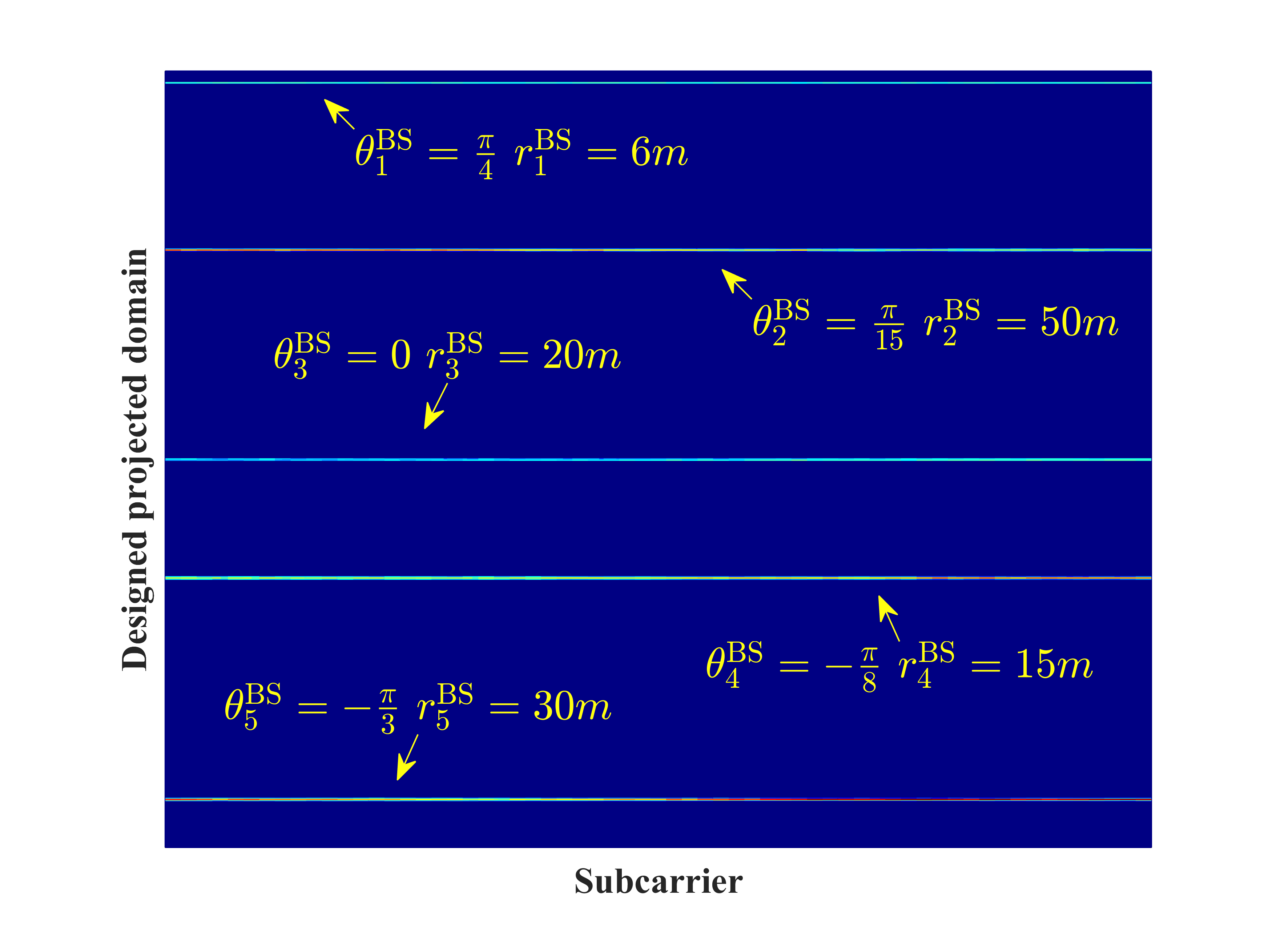}
		\label{fig3:polar domain}
	}	
	\subfigure[The piece-wise planar wave approximation of LoS component.]{	
	\includegraphics[width=0.62\columnwidth, keepaspectratio]{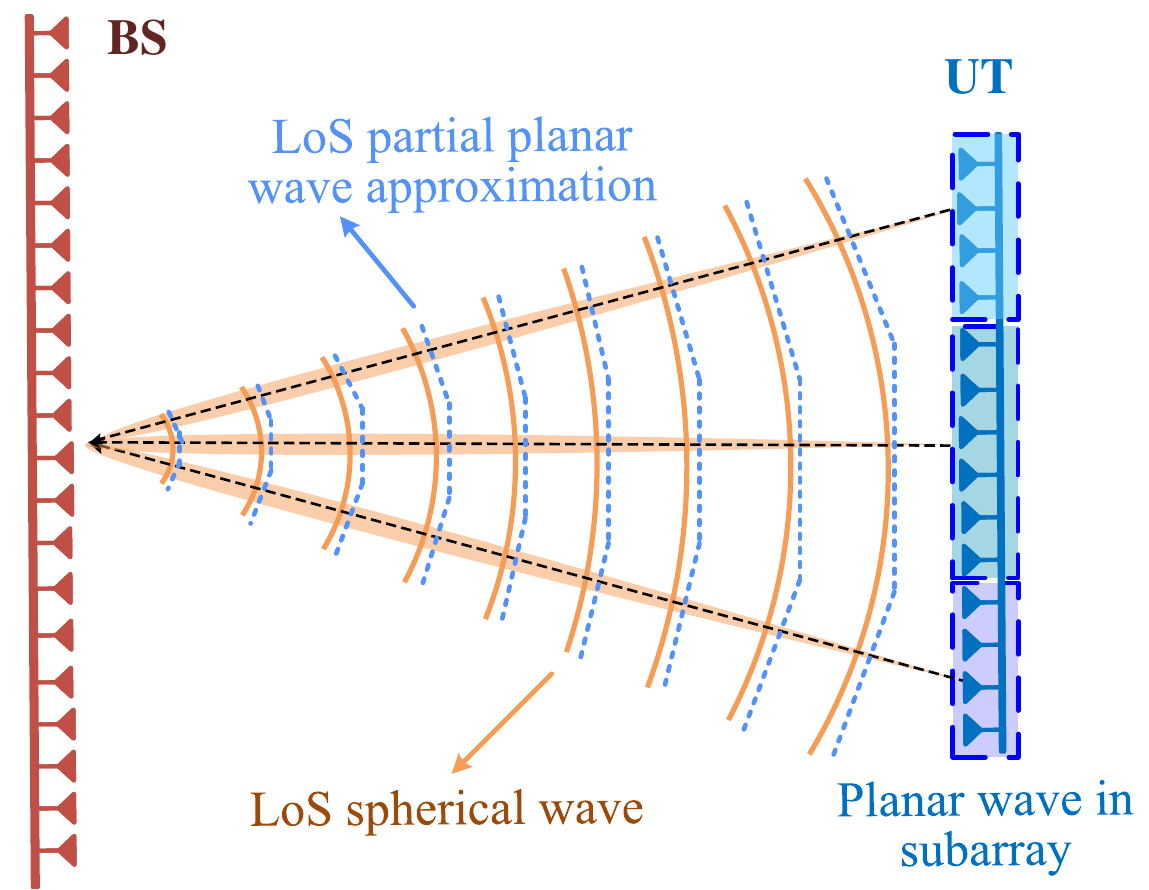}
	\label{fig3:los pca}
    }	
	\captionsetup{font={color = {black}}, justification = raggedright,labelsep=period}
	\caption{The sparse representation of HFBS channel, where the system parameters are the carrier frequency $f_c = 47\rm GHz$, system bandwidth $\text{BW}=5\text{GHz}$, $N_{\rm BS} = 512$, $N_{\rm UT} = 32$, and $M = 64$ for ease of illustration. }	
	\vspace{-5mm}	
\end{figure*}


In the case of hybrid far- and near-field propagation effect, the planar wave approximation based projection matrix fails to exactly capture the inherent characteristics of the channel, 
leading to a degradation in the sparsity of the channels represented in such a projection matrix, as illustrated in Fig. \ref{fig3:dft}. 
To address this issue, a more accurate Fresnel approximation \cite{Fresnel} is developed for the distance $r_{l,n_1}^{\rm BS}$ based on a second-order Taylor expansion as
\begin{align} \label{eq:Fresnel approximation}
	r_{l,n_1}^{\rm BS} = \sqrt{\left( {r_l^{\rm BS}} \right)^2 + \delta^2_{n_1} d^2 - 2 r_l^{\rm BS} \delta_{n_1} d \sin(\theta_l^{\rm BS}) } \notag \\
	\approx  r_{l}^{\rm BS} - \delta_{n_1} d \sin(\theta_l^{\rm BS}) + \frac{\delta_{n_1}^2 d^2 \cos^2 \theta_l^{\rm BS}}{2{r_l^{\rm BS}}}.
\end{align}
Based on this approximation, the BS array manifold of each MPC depends on not only its AoD/AoA but also on the distance between the BS and scatterers.  
To this end, \cite{cmy,lzr} proposed a non-uniform two-dimensional sampling lattice that considers both angles and distances as
\begin{subequations} \label{eq:sampling grid}
\begin{align}
	 & \varTheta = \left\{ \sin \theta_n = \frac{2n-\rho N_{\rm BS}+1}{\rho N_{\rm BS}} \Big| n=0,1,\dots, \rho N_{\rm BS}-1, \forall s \right\},  
\end{align}	 
\begin{align}
	 & \mathcal{R} = \left\{  \frac{1}{r_{n,s}} = \frac{s \eta}{4R_r \cos^2\theta_n} \Big| s=0,1,\dots, S-1 , \forall n  \right\}, 
\end{align}
\end{subequations}
where $\rho N_{\rm BS}$ and $S$ are the number of grids along the angles and distances, respectively,   
and $R_r = 2D_{\rm BS}^2 / \lambda_c$ is the Rayleigh distance corresponding to the antenna array aperture $D_{\rm BS}$. Furthermore, $\eta$ is a scaling factor utilized to control the sampling density.
     
In this way, the projection matrix $\boldsymbol{\Phi} [m] \in \mathbb{C}^{N_{\rm BS} \times \rho N_{\rm BS}S}$ can be designed as 
\begin{align}\label{eq:polar-domain projetion matrix}
	\boldsymbol{\Phi} [m] = \left\{ \mathbf{a}_{\rm BS} (\theta,r,m) | \theta \in \varTheta, r \in \mathcal{R} , \forall \theta, r \right\},
\end{align}
in which the frequency-selective array manifold $\mathbf{a}_{\rm BS} (\theta,r,m)$ sampled from the defined 2D lattice forms the $f(\theta,r)$-th candidate columns of the projection matrix $\boldsymbol{\Phi} [m]$.
It is essential to note that when $s=0$, $r_{n,s} \rightarrow \infty$, the defined sampling grids match the traditional virtual-angular lattices in the far-field. This alignment indicates  that the projection matrix $\boldsymbol{\Phi} [m]$ serves as a generalized representation of channels in both far-field and near-field.
Furthermore, in contrast to the frequency-flat projection matrix leading to the offset of sparse support sets across different subcarriers \cite{cmy}, the proposed matrix 
considers the frequency-dependent behavior of the array manifold to ensure an identical 
support set across the entire frequency spectrum, as illustrated in Fig. \ref{fig3:polar domain}.

As a result, the NLoS components of the HFBS channel can be readily represented in a sparse manner using the defined projection matrix. Furthermore, while the LoS component in \eqref{eq:los} cannot be straightforwardly decomposed into the product of two array manifold vectors, it can be approximated as follows using the piece-wise subarray planar wave for UT's array as depicted in Fig. \ref{fig3:los pca} 
\begin{align}
	\mathbf{H}_{\rm UL}^{\rm LoS}[m]  \approx  \sqrt{\beta_{\rm LoS}} \left[ {\mathbf{a}_{\rm BS}^{(1)}} (m)  {\mathbf{a}_{\rm UT}^{(1)}}^{\rm H} (m), \dots,{\mathbf{a}_{\rm BS}^{(G)}}(m) {\mathbf{a}_{\rm UT}^{(G)}}^{\rm H}(m)  \right],
\end{align}
where $G$ indicates the number of subarrays divided for UT's array, ${\mathbf{a}_{\rm UT}^{(g)}}(m) = \mathbf{a}_{\rm UT} (\theta_g^{\rm UT},\infty,m)$ and ${\mathbf{a}_{\rm BS}^{(g)}} (m) = (\theta_g^{\rm BS},r_g^{\rm BS},m)$ denote the UT's array steering vectors of the $g$-th subarray and the BS's array steering vectors, respecitvely.
Under this approximation, the LoS component can be represented sparsely with $G$ components as   
\begin{align}\label{eq:los app}
	 & \mathbf{H}_{\rm UL}^{\rm LoS}[m]  \approx \sqrt{\beta_{\rm LoS}}   \notag \\ &  \times \sum_{g=1}^{G} 
	{\mathbf{a}_{\rm BS}^{(g)}} (m) \left[ {\mathbf{0}_{(g-1)N_{\rm UT}/G}, \,\mathbf{a}_{\rm UT}^{(g)}} (m), \, \mathbf{0}_{N_{\rm UT}-gN_{\rm UT}/G}  \right]^{\rm H}. 
\end{align}
Therefore, the channel estimation in \eqref{eq:eff CE} is equivalent to a structured sparse signal recovery problem 
\begin{align} \label{eq:final CE}
	{\mathbf{Y}}_{\rm UL}[m] = \mathbf{A} [m] \mathbf{H}_{\rm UL}^{P}[m]  + {\mathbf{N}}[m], \,  \forall m, 
\end{align} 
where $\mathbf{A} [m] = \mathbf{W}^{\rm H}_{\rm UL} \boldsymbol{\Phi} [m] \in  \mathbb{C}^{P \times \rho N_{\rm BS} S}$ is the effective sensing matrix and $\mathbf{H}_{\rm UL}^{P}[m] \in \mathbb{C}^{\rho N_{\rm BS} S \times N_{\rm UT}}$ denotes the high-dimensional projected sparse channel matrix.

\subsection{AMP-EM-Based Channel Estimation} \label{s3.3}

It is well-known that AMP is an efficient iterative Bayesian algorithm for CS problems, which employs low-complexity message passing for approximating the traditional minimum mean square error (MMSE) estimation \cite{amp}.
For notation brevity, the superscript, subscript, and index $m$ have been omitted for ${\mathbf{Y}}_{\rm UL}[m]$ and $\mathbf{H}_{\rm UL}^{P}[m]$ hereinafter, and the different entries of matrix is distinguished by its subscript, e.g., $h_{k,n} = \left[ \mathbf{H} \right]_{k,n}$. 
Starting with the MMSE estimate of $\mathbf{H}$, we can obtain its posterior mean as 
\begin{align}
	\hat{h}_{k,n} = \mathbb{E}\{{h}_{k,n} | \mathbf{Y}\}  = \int {h}_{k,n} p({h}_{k,n}|\mathbf{Y}) \, {\rm d} {h}_{k,n}, \forall k,n,  
\end{align}
where $p({h}_{k,n}|\mathbf{Y})$ denotes the marginal posterior distribution expressed as follows:
\begin{align} \label{eq:marginal posterior distribution}
	p({h}_{k,n}|\mathbf{Y}) = \int  p(\mathbf{H}|\mathbf{Y}) \, {\rm d} 
	\mathbf{H}_{\sim\{k,n\}}.
\end{align}
\eqref{eq:marginal posterior distribution} requires the high-dimensional
integral of joint posterior distribution $p(\mathbf{H}|\mathbf{Y})$ w.r.t. each variable entry of $\mathbf{H}$ except ${h}_{k,n}$. 
Based on the Bayesian theorem, $p(\mathbf{H}|\mathbf{Y})$ can be represented as
\begin{align}
	p(\mathbf{H}|\mathbf{Y}) & = \frac{p(\mathbf{Y}|\mathbf{H}) \, p_0(\mathbf{H})}{p(\mathbf{Y})} \notag \\
	& = \frac{1}{p(\mathbf{Y})}	
	\prod_{n=1}^{N_{\rm UT}} \left[  \prod_{p=1}^{P}  p(y_{p,n} | \mathbf{H} ) \prod_{k=1}^{\rho N_{\rm BS} S} \hspace{-4pt} p_0(h_{k,n})  \right].
\end{align}
Under the assumption of AWGN, the above likelihood function $p(\mathbf{Y}|\mathbf{H})$ obeys complex Gaussian distribution 
\begin{align}
	p(y_{p,n} | \mathbf{H} ) = -\frac{1}{\pi \sigma^2} {\rm exp} \left( -\frac{1}{\sigma^2} \left| y_{p,n} - \sum\nolimits_{k} a_{p,k} h_{k,n}  \right|^2  \right).
\end{align} 
By following the classic AMP literature \cite{mxm}, we consider the prior distribution of $\mathbf{H}$ to follow an i.i.d. Bernoulli-Gaussian distribution for effectively capturing 
its sparsity, i.e., 
\begin{align}\label{eq:prior}
	p_0 (h_{k,n}) = (1-\lambda_{k,n}) \delta(h_{k,n}) + \lambda_{k,n} \mathcal{CN}(h_{k,n};\mu_{k,n},\gamma_{k,n}),
\end{align} 
where $0 \le \lambda_{k,n} \le 1$ denotes the non-zero probability of $h_{k,n}$, and $\mu_{k,n}$, $\gamma_{k,n}$ are the prior mean and variance of a Gaussian distribution, respectively. 

The joint distribution $p(\mathbf{H},\mathbf{Y})$ can be represented by a bipartite
graph $\mathcal{G} = (\mathcal{V},\mathcal{F},\mathcal{E})$, which consists of variable nodes $\mathcal{V}$, factor nodes $\mathcal{F}$, and the corresponding edges $\mathcal{E}$ \cite{sum-product}.
This suggests the use of message passing algorithms contributes to rigorous solution of the marginal posterior distribution in \eqref{eq:marginal posterior distribution}, and the messages approximation can further reduce the complexity of high-dimensional integral.
Therefore, it yields \cite{kml}
\begin{align} \label{eq:approx}
	p({h}_{k,n}|\mathbf{Y}) \propto p_0({h}_{k,n}) \, \mathcal{CN}(h_{k,n};R_{k,n}^t,\Sigma_{k,n}^t),
\end{align}
where $t$ indicates the iteration index, $\Sigma_{k,n}^t$ and $R_{k,n}^t$ are the message parameters updated at variable nodes $\mathcal{V}$ as 
\begin{subequations}\label{eq:variable node}
\begin{align} 
	 \Sigma_{k,n}^t & = \left[ \sum\nolimits_p \frac{|a_{p,k}|^2}{\sigma^2 + V_{p,n}^t} \right]^{-1},\\
	 R_{k,n}^t & = \hat{h}_{k,n}^t + \Sigma_{k,n}^t \sum\nolimits_{p} \frac{a_{p,k}^* (y_{p,n} - Z_{p,n}^t)}{\sigma^2 + V_{p,n}^t},
	\end{align}
\end{subequations}
while $V_{p,n}^t$ and $Z_{p,n}^t$ are the message parameters updated at factor nodes $\mathcal{F}$ as
\begin{subequations} \label{eq:factor node}
\begin{align}
	& V_{p,n}^t = \sum\nolimits_{k} |a_{p,k}|^2 \hat{v}_{k,n}^t, \\
	& Z_{p,n}^t = \sum\nolimits_{k} a_{p,k} \hat{h}_{k,n}^t - \frac{V_{p,n}^t}{\sigma^2 + V_{p,n}^{t-1}} (y_{p,n} - Z_{p,n}^{t-1}).
\end{align}
\end{subequations}
where $\hat{h}_{k,n}^t$ and $\hat{v}_{k,n}^t$ are defined in \eqref{eq:posterior}.
By combining \eqref{eq:prior} and \eqref{eq:approx}, the marginal posterior distribution can be merged as 
\begin{align}  \label{eq:posterior pdf}
	p({h}_{k,n}|\mathbf{Y}) = (1-\pi_{k,n}^t) \, \delta ({h}_{k,n}) + \pi_{k,n}^t \, \mathcal{CN} (h_{k,n};\xi_{k,n}^t,\zeta_{k,n}^t),
\end{align}
where 
\begin{subequations}
\begin{align}
	& \xi_{k,n}^t = \frac{\Sigma_{k,n}^t \mu_{k,n} +R_{k,n}^t\gamma_{k,n}}{\Sigma_{k,n}^t+\gamma_{k,n}},\quad \zeta_{k,n}^t = \frac{\Sigma_{k,n}^t \gamma_{k,n}}{\Sigma_{k,n}^t+\gamma_{k,n}}, \\
	& \pi_{k,n}^t = \frac{\lambda_{k,n}}{\lambda_{k,n}+(1-\lambda_{k,n}) \, {\rm exp} (-\mathcal{L}^t_{k,n})}, \\
	& \mathcal{L}^t_{k,n} = \frac{1}{2} \ln \frac{\Sigma_{k,n}^t}{\Sigma_{k,n}^t+\gamma_{k,n}} + \frac{\left|R_{k,n}^t\right|^2}{2\Sigma_{k,n}^t} - \frac{\left| R_{k,n}^t- \mu_{k,n} \right|^2}{2(\Sigma_{k,n}^t+\gamma_{k,n})}.
\end{align}
\end{subequations}
Then, the posterior mean and variance can be explicitly calculated as
\begin{subequations}\label{eq:posterior}
\begin{align}
	& \hat{h}_{k,n}^{t+1} = \mathbb{E}\{{h}_{k,n} | R_{k,n}^t, \Sigma_{k,n}^t\} = \pi_{k,n}^t \xi_{k,n}^t, \\
	& \hat{v}_{k,n}^{t+1} = \mathbb{V} \{{h}_{k,n} | R_{k,n}^t, \Sigma_{k,n}^t\} = \pi_{k,n}^t \left( |\xi_{k,n}^t|^2 + \zeta_{k,n}^t \right) - |\hat{h}_{k,n}^{t+1}|^2.
\end{align}
\end{subequations}

The message switching procedures mentioned in \eqref{eq:variable node}-\eqref{eq:posterior} constitute the fundamental steps of the AMP algorithm. However, several concerns arise when applying the AMP algorithm to practical systems:
(a) The hyper-parameters in the prior assumption and the noise variance are often unknown in practice.
(b) The column-wise independent solution to the channel estimation problem neglects the common support structure that arises from spatial and frequency correlations.
(c) The AMP algorithm is primarily designed for large system limits and typically assumes the use of i.i.d. Gaussian sensing matrices. This assumption is unrealistic for the problem described in \eqref{eq:final CE}.

\begin{algorithm}[t] 	
	\caption{AMP-EM-Based HFBS Channel Estimation}
	\label{alg:amp}
	\LinesNumbered 
	\KwIn{Received reference signal ${\mathbf{Y}}_{\rm UL}[m]$, sensing matrix $\mathbf{A} [m]$, maximum iteration number $T_{\rm iter}$, damping parameter $\epsilon$.}
	\KwOut{Estimated channel matrix $\widehat{\mathbf{H}}_{\rm UL}[m]$, and the average noise variance $\overline{\hat{\sigma}^2} [m]$.}
	Initialize hyper-parameters $\boldsymbol{\theta}^1$ as \cite{init} recommended in case
	of trapping in local extremum, and initialize other parameters $V_{p,n}^0[m] = 1$,  $Z_{p,n}^0[m] = y_{p,n}$, $\hat{h}_{k,n}^{1} [m] = \mu_{k,n}^1 [m]$, $\hat{v}_{k,n}^{1}[m] = \gamma_{k,n}^{1} [m]$. Set iteration index $t=1$. 
	
	\For{$1 \le t \le T_{\rm iter}$}{		
		\For{$1 \le m \le M$}{
			Update factor nodes $V_{p,n}^t [m]$, $Z_{p,n}^t [m]$ in line with  \eqref{eq:factor node} and implement the damping process:
			$V_{p,n}^t [m] \leftarrow \epsilon V_{p,n}^{t-1} [m]+ (1-\epsilon) V_{p,n}^{t} [m]$, 
			$Z_{p,n}^t [m] \leftarrow \epsilon Z_{p,n}^{t-1} [m] + (1-\epsilon) Z_{p,n}^{t} [m]$.
			
			Update variable nodes $\Sigma_{k,n}^t [m]$,$R_{k,n}^t [m]$ according to  \eqref{eq:variable node}, and the posterior mean and variance $\hat{h}_{k,n}^{t+1} [m]$, $\hat{v}_{k,n}^{t+1} [m]$ according to  \eqref{eq:posterior}.
		}
		Learn hyper parameters $\lambda_{k,n}^{t+1} [m]$, $\mu_{k,n}^{t+1} [m]$, $\gamma_{k,n}^{t+1} [m]$, and $(\sigma^2_{p,n} )^{t+1}[m]$ in line with rules in \eqref{eq:lambda}, \eqref{eq:hyper parameters}.
		
		$t \leftarrow t+1$. 
	}
	
	\KwResult{Return the estimated channel $\widehat{\mathbf{H}}_{\rm UL}[m] =  \boldsymbol{\Phi} [m] \widehat{\mathbf{H}}_{\rm UL}^{P}[m]$, and the average
		 noise variance $\overline{\hat{\sigma}^2} [m] = \frac{1}{P N_{\rm UT}}\sum_{p,n} (\sigma^2_{p,n} )^{T_{\rm iter}+1}[m]$.}
\end{algorithm}

To overcome the first challenge, EM can be exploited to learn the unknown hyper-parameters $\boldsymbol{\theta}=\{\lambda_{k,n},\mu_{k,n},\gamma_{k,n},\sigma^2, \forall k,n\}$ as \cite{em}
\begin{subequations}
\begin{align}
	  Q(\boldsymbol{\theta},\boldsymbol{\theta}^{t}) & = \mathbb{E} 
	\left\{ \ln p(\mathbf{H},\mathbf{Y})  | \mathbf{Y}; \boldsymbol{\theta}^{t} \right\}, \\
	 \quad\quad  \boldsymbol{\theta}^{t+1} &  = \mathop{\arg\max}\limits_{\boldsymbol{\theta}} \, Q(\boldsymbol{\theta},\boldsymbol{\theta}^{t}), \label{eq:m-step}
\end{align}  
\end{subequations}
where $\mathbb{E}\left\{ \cdot| \mathbf{Y}; \boldsymbol{\theta}^{t} \right\}$ denotes the 
expectation with respect to the posterior distribution $p(\mathbf{H}|\mathbf{Y}; \boldsymbol{\theta}^{t})$, and its approximation is concluded in \eqref{eq:posterior pdf}.
By updating one of the elements in \eqref{eq:m-step} at a time and holding others constant, the learning rules of the hyper-parameters can be summarized as \cite{kml}
\begin{subequations}  \label{eq:hyper parameters}
\begin{align}\label{eq:EM sparsity} 
	& \lambda_{k,n}^{t+1} = \pi_{k,n}^t, \quad \mu_{k,n}^{t+1} = \frac{\sum_k \pi_{k,n}^t \xi_{k,n}^t}{\sum_k \pi_{k,n}^t}, 
\end{align}
\begin{align}
	& \gamma_{k,n}^{t+1} = \frac{\sum_k \pi_{k,n}^t (|\mu_{k,n}^{t} - \xi_{k,n}^t|^2 + \zeta_{k,n}^t) }{\sum_k \pi_{k,n}^t}, 
\end{align}
\begin{align}
	& (\sigma^2_{p,n})^{t+1} = \frac{1}{P} \sum\nolimits_{p}
	\left( \frac{|y_{p,n}-Z_{p,n}^t|^2}{\left| 1+V_{p,n}^t/(\sigma^2_{k,n})^{t} \right|^2} + \frac{(\sigma^2_{k,n})^{t} V_{p,n}^t}{(\sigma^2_{k,n})^{t}+V_{p,n}^t} \right).
\end{align}
\end{subequations}
To further address the second concern aforementioned, it is reasonable to assign a common non-zero probability $\lambda_{k,n}$ for each different columns and subcarriers of $\mathbf{H}$, which indicates that the learning rule in \eqref{eq:EM sparsity} and can be refined as 
\begin{align} \label{eq:lambda}
	\lambda_{k,1}^{t+1} [1] = \cdots = \lambda_{k,N_{\rm UT}}^{t+1} [M] = \frac{1}{MN_{\rm UT}} \sum_{n,m} \lambda_{k,n}^{t+1} [m].
\end{align}  

{ Moreover, the damping operation \cite{damping} could be introduced to prevent AMP from divergence due to the nonideal sensing matrix $\mathbf{A} [m]$ and the medium size problem mentioned in the concern (c).}
To sum up, the overall procedure of the AMP-EM-based HFBS channel estimation algorithm is presented in Alg. \ref{alg:amp}.

\section{Location Sensing}

This section leverages the acquired CSI to achieve simultaneous location sensing of the scatterers and UTs without requiring additional radio resources.
Specifically, we first extract the targets' angle and distance information from the spherical electromagnetic wavefront based on the channel estimation results to sense the scatterers' locations. 	
Then, treating these sensed scatterers as VAs, the UT's location can be coarsely estimated according to the propagation geometry. 
To further enhance the sensing performance, we utilize the estimated relative delay of MPCs to refine the UT's and scatterers' location through an iterative procedure.

\subsection{VAs Location Mapping} \label{sec:4.1}

Under the spherical wave assumption, the HFBS channel depends not only on the angle of the target, but also on its distance.
Therefore, it is feasible to simultaneously derive the angle and distance features of the target from the channel state information.
According to the analysis in Section \ref{sec:Projection Matrix Design}, under ideal conditions, the non-zero support set of sparse channel $\mathbf{H}_{\rm UL}^{P}[m]$ exactly reflects the approximate LoS components and the NLoS MPCs of the HFBS channel, and their angle and distance parameters are associated with the spatial spatial sampling grid defined in \eqref{eq:sampling grid}. 
In this way, we can map the location of the center of UT's subarrays and the scatterers base on the channel estimation results. 

One of the premise to achieve location mapping lies on accurate knowleadge about MPCs' number.
To this end, we estimate the effective MPCs' number $L^{\prime} = G + L$ based on the minimum description length (MDL) principle,
which selects the parametric probability model that fits the dataset best based on the maximum likelihood criterion, so as to estimate the number of signal sources without the need of empirical threshold \cite{mdl}.

Firstly, in order to overcome the rank-deficient problem of covariance matrix caused by coherent multipath signals, the covariance matrix of received pilot signals is calculated with the spatial smoothing technique \cite{sptial-smoothing} as  
\begin{align}\label{eq:auto-correlation}
	\mathbf{R}_{\rm yy}^K [m]  & = \frac{1}{K+1} \sum_{i=0}^K \mathbf{I}_{N_{\rm UT}-K,i} \mathbf{R}_{\rm yy} [m] \mathbf{I}_{N_{\rm UT}-K,i}^{\rm T},
\end{align}
where $\mathbf{R}_{\rm yy} [m] \approx {\mathbf{Y}}_{\rm UL}^{\rm H} [m] {\mathbf{Y}}_{\rm UL}[m]$ is the estimated covariance matrix of the original signal, while $\mathbf{I}_{N_{\rm UT}-K,i} = \left[ \mathbf{0}_{(N_{\rm UT}-K) \times i} \; \mathbf{I}_{N_{\rm UT}-K} \; \mathbf{0}_{(N_{\rm UT}-K) \times (K-i)}  \right] \in  \mathbb{C}^{(N_{\rm UT}-K) \times N_{\rm UT}}$.
By the spectrum decomposition theorem, if there are $L^{\star} $ of the multipath, $\mathbf{R}_{\rm yy}^K$ can be decomposed into
\begin{align}
	\mathbf{R}_{\rm yy}^K = \sum_{i=1}^{L^{\star}} (\lambda_i-\sigma^2_n) \mathbf{u}_i \mathbf{u}_i^{\rm H} + \sigma^2_n \mathbf{I}_{N_{\rm UT}-K},
\end{align}
where $\lambda_i$ and $\mathbf{u}_i$ are the eigenvalues and eigenvectors of $\mathbf{R}_{\rm yy}^K$, respectively.
Defines a set of parameters for the received signal model$\boldsymbol{\eta} = \left\{  \lambda_1,\dots,\lambda_{L^{\star}}, \mathbf{u}_1,\dots,\mathbf{u}_{L^{\star}},  \sigma^2_n \right\}$, the estimate of $L^{\star}$ can be obtained based on the MDL rule as 
\begin{align}\label{eq:L estimate}
	\hat{L}^{\prime} = - \mathop{\arg\min}\limits_{L^{\star}} \ \left[ \log P({\mathbf{Y}}_{\rm UL}|{\boldsymbol{\eta}}) + \gamma(L^{\star}) \right],
\end{align}
where $P({\mathbf{Y}}_{\rm UL}|{\boldsymbol{\eta}})$ is the joint probability density distribution of the received pilot signals, while $\gamma(L^{\star}) = \frac{1}{2}L^{\star}(2N_{\rm UT}-L^{\star}-K)\log N_{\rm BS}$.
Under the assumption of complex Gaussian distribution, it can be further expressed as \eqref{eq:mdl}, which is shown at the top of the next page, where
$\hat{\lambda}_{i}$ is the estimate of eigenvalue by decomposing the estimated covariance matrix $\mathbf{R}_{\rm yy}^K$.
By calculating the covariance matrix of the received pilot signal on different subcarriers, $M$ independent estimation results can be obtained in parallel. 
Finally, the fusion of different estimation results can be completed by the mode operation.

\begin{figure*}[t]	
	\begin{align}\label{eq:mdl}
		\hat{L}^{\prime} =\mathop{\arg\min}\limits_{L^{\star}} \ \left\{N_{\rm BS} (N_{\rm UT}-K-L^{\star}) \log \left[\frac{\frac{1}{N_{\rm UT}-K-L^{\star}} \sum_{i=L^{\star}+1}^{N_{\rm UT}} \hat{\lambda}_{i}}{\left(\prod_{i=L^{\star}+1}^{N_{\rm UT}} \hat{\lambda}_{i}\right)^{\frac{1}{N_{\rm UT}-K-L^{\star}}}}\right]+\gamma(L^{\star}) \right\},
	\end{align}
	\hrulefill
\end{figure*}

\begin{figure}[t]	
	\centering
	\includegraphics[width=0.8\columnwidth, keepaspectratio]{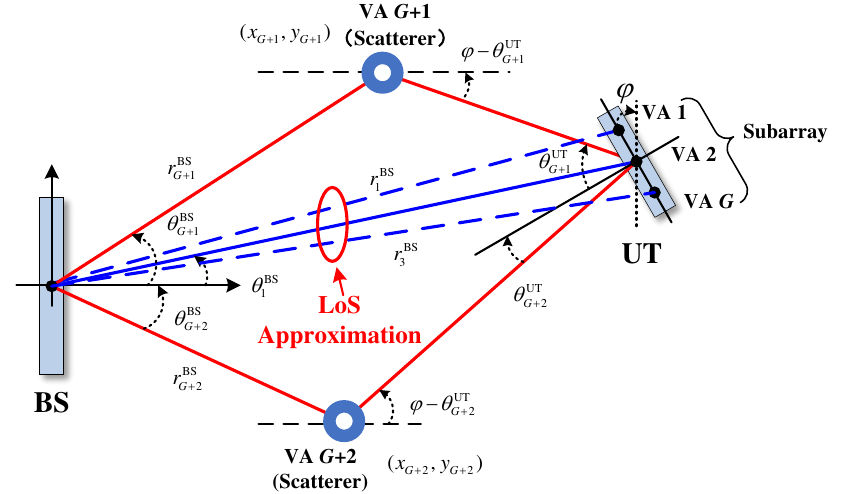}
	\captionsetup{font={color = {black}}, justification = raggedright,labelsep=period}
	\caption{Propagation geometry of the location sensing problem in the UM-MIMO system.}
	\label{fig4:geometric model}	
	\vspace{-5mm}
\end{figure}

When the corresponding row energy of the estimated sparse channel is greater than the noise threshold, the MPCs can be considered to exist. 
Extract the angle and distance parameters of them
\begin{align}\label{eq:eff grid}
	(\tilde{\varTheta},\tilde{\mathcal{R}}) = \mathop{\arg}\limits_{\theta \in {\varTheta}, r \in {\mathcal{R}}} \left\{ \widehat{h}_{\rm UL}^{P}(\theta,r) > N_{\rm UT} \sum_{m}\overline{\hat{\sigma}^2} [m]  \right\},
\end{align}
where $	\hat{h}_{\rm UL}^{P}(\theta,r) = \sum_{n=1}^{N_{\rm UT}} \sum_{m=1}^{M} \left|\left\{  \widehat{\mathbf{H}}_{\rm UL}^{P} [m] \right\}_{f(\theta,r),n} \right|^2$.
Unfortunately, due to the finite-resolution sampling lattice, the energy leakage is an  inevitable negative effect.
As a result, the grids set $(\tilde{\varTheta},\tilde{\mathcal{R}})$ exhibits a cluster distribution, where $\hat{L}^{\prime}$ truthful peaks are surrouded by a few ghost points.
In order to divide clusters before peak identification, K-Means clustering algorithm \cite{k-means} can be exploited to select the spatial grids subset $(\tilde{\varTheta}_l,\tilde{\mathcal{R}}_l)$ for the $l \, (1 \le l \le \hat{L}^{\prime})$-th effective MPC. 
In this case, the angle and distance of $l$-th MPC corresponding to the summit of the $l$-th cluster grids can be represented as 
\begin{align} \label{eq:grid match}
	(\hat{\theta}_l^{\rm BS},\hat{r}_l^{\rm BS})  = \mathop{\arg\max}\limits_{\theta \in \tilde{\varTheta}_l, r \in \tilde{\mathcal{R}}_l} \, \widehat{h}_{\rm UL}^{P}(\theta,r).
\end{align}

Considering the location sensing geometry  illustrated in Fig. \ref{fig4:geometric model}, where the BS is located at the origin of coordinates, we can coarsely map the position of VAs 
with the orientation and distances measured above as $(\hat{x}_l,\hat{y}_l) = (\hat{r}_l^{\rm BS} \cos \hat{\theta}_l^{\rm BS},\hat{r}_l^{\rm BS} \sin \hat{\theta}_l^{\rm BS}  )$,
where we treat the sensed UT subarray centers and scatterers equally as VAs, and they can contribute to UT location sensing hereinafter. 
In order to distinguish the sensed UT subarray centers from the scatterers, we adopt the following method: VAs satisfying the array aperture constraint are inferred as UT subarray centers, and their coordinate set $\mathcal{G} = \{ l | 1 \leq l \leq \hat{L^{\prime}}, (\hat{x}_l-\hat{x}^{\star})^2 + (\hat{y}_l-\hat{y}^{\star})^2 \leq D_{\rm UT} \}$, where $(\hat{x}^{\star},\hat{y}^{\star})$ is the VAs with the maximum energy and thus can be truthfully inferred as LoS components, while $D_{\rm UT}$ is the antenna array aperture of UT. The remains are classified as scatterers and their coordinate set $\mathcal{S} = \{ l | 1 \leq l \leq \hat{L}^{\prime}, l \notin \mathcal{G} \}$.
 
\subsection{Coarse UT Location Sensing}

In order to acquire the UT's position with the aid of sensed VAs, one of the prerequisites lies in acquiring the orientation between the UT and VAs. 
For this purpose, motivated by the channel eigenvector decomposition in 
\eqref{eq:channel model} and LoS approximation in \eqref{eq:los app},
and the BS steering vector can be estimated as $\mathbf{a}_{\rm BS} (\theta_l^{\rm BS},r_l^{\rm BS},m) \approx \mathbf{a}_{\rm BS} (\hat{\theta}_l^{\rm BS},\hat{r}_l^{\rm BS},m)$, thus the angle and distance can be estimated via matching pursuit as  
\begin{align} \label{eq:power specturm orientation}
	& (\hat{\theta}_l^{\rm UT},\hat{r}_l^{\rm UT}) = \notag \\
	& \mathop{\arg\max}\limits_{\theta,r}
	\mathop{\max}\limits_{m}
	{\left\Vert  \left\{  \widehat{\mathbf{H}}_{\rm UL}^{P} [m] \right\}_{\tiny f(\hat{\theta}_l^{\rm BS},\hat{r}_l^{\rm BS}),:}
		\, \mathbf{a}_{\rm UT}^* (\theta,r,m)  \right\Vert^2}.   
\end{align}
It is noteworthy that since the array aperture of UT is much smaller than that of BS, the Rayleigh distance observed from UT could be relatively small so that the uncertainty of  $\hat{r}_l^{\rm UT}$ aggravates. 
In this context, the geometric propagation model used for UT's location sensing in Fig. \ref{fig4:geometric model} can be represented irrespective of ${r}_l^{\rm UT}$ as
\begin{align}\label{eq:geometric model}
	\begin{cases}
		\varphi = \theta_l^{\rm BS} + \theta_l^{\rm UT}, l \in \mathcal{G}, \\
		(y_l-y)/(x_l - x) = \tan \, (\varphi - \theta_l^{\rm UT}), l \in \mathcal{S}, \\
		(y_l-y)/(x_l - x) = \tan \, (\varphi - \pi/2), l \in \mathcal{G},
	\end{cases}
\end{align}
where $(x,y)$ is the location of UT to be determined, and $\varphi$ is the inclination
angle of UT's array. 
Furthermore, it can be rewritten with a matrix form as 
\begin{align}\label{eq:los localization}
	\underbrace{
		\begin{bmatrix}
			{y}_{l \in \mathcal{S}} - \tan ({\varphi} -\theta_{l \in \mathcal{S}}^{\rm UT} ) {x}_{l \in \mathcal{S}}    \\
			\vdots 	\\
			{y}_{l \in \mathcal{G}} - \tan ({\varphi} - \pi/2 ) {x}_{l \in \mathcal{G}}
		\end{bmatrix}
	}_{\mathcal{Y} \in \mathbb{R}^{{L^{\prime}} \times 1}}
	=
	\underbrace{\begin{bmatrix}
			-\tan ({\varphi} - {\theta}_{l \in \mathcal{S}}^{\rm UT}) & 1   \\
			\vdots & \vdots	\\
			-\tan ({\varphi} - \pi/2) & 1
		\end{bmatrix}
	}_{\mathcal{H} \in \mathbb{R}^{{L^{\prime}} \times 2}}
	\begin{bmatrix}
		{x} \\ {y}	
	\end{bmatrix}.
\end{align}                      
In view of the non-uniform sampling pattern defined in \eqref{eq:sampling grid}, 
the further the VAs locate, the sparser the corresponding sampling grid is. 
If the scatterer falls into a region with sparser spatial sampling grids, the location sensing error for those VAs would aggravate, and their reliability for UT localization would decline. 
Therefore, one viable solution to the above problem is the weighted least squares (w-LS) method \cite{w-ls},
which introduces a weighted matrix assigning higher weights to VAs with smaller sampling interval
and lower weights to VAs with larger sampling interval, thereby enhancing the accuracy of the estimation, i.e.,
\begin{align}\label{eq:wls}
	\left[  \hat{x} \;  \hat{y} \right]^{\rm T}	
	=   
	\left( {\hat{\mathcal{H}}^{\rm T}} \mathbf{W} \hat{\mathcal{H}} \right)^{\rm -1} \hat{\mathcal{H}}^{\rm T} \mathbf{W} \hat{\mathcal{Y}}, 
\end{align}
where $\hat{\mathcal{Y}}$ and $\hat{\mathcal{H}}$ are estimated by replacing the corresponding element with the sensing results in \eqref{eq:grid match} respectively, while $\hat{\varphi}$ can be estimated as $\hat{\varphi} = \frac{1}{|\mathcal{G}|_c} \sum_{l \in \mathcal{G}} \left( \theta_l^{\rm BS} + \theta_l^{\rm UT} \right)$. Moreover, the weighted matrix can be expressed as $\mathbf{W} = {\rm diag} \left\{ \frac{1}{|\hat{r}_1^{\rm BS}-\mathcal{N}(\hat{\theta}_1^{\rm BS},\hat{r}_1^{\rm BS})|^2}, \dots,  \frac{1}{|\hat{r}_{\hat{L^{\prime}}}^{\rm BS}-\mathcal{N}(\hat{\theta}_{\hat{L}^{\prime}}^{\rm BS},\hat{r}_{\hat{L}^{\prime}}^{\rm BS})|^2} \right\}$, and $\mathcal{N} (\theta,r)$ denotes the distance parameter of nearest adjacent grid set in \eqref{eq:sampling grid} corresponding to $(\theta,r)$.

\begin{remark}
	In fact, the above mentioned UT localization method can also be extended to blocked LoS case.
	In the absence of LoS component, compared with \eqref{eq:geometric model}, the only difference for location sensing problem lies in the absence of information about LoS orientation $(\theta_1^{\rm BS}$, $\theta_1^{\rm UT})$, which does not yet affect its rigorous LS solution when the number of sensed VAs meets $\hat{L}^{\prime} \ge 3$.  	     
\end{remark}

\subsection{Sensing Result Refinement}

The aforementioned coarse UT's location sensing result only depends on the angle and distance information extracted from the spherical wavefront, which applies to both wideband and narrowband sensing tasks in sharp contrast to the traditional TDoA-based methods \cite{w-ls}.  
On that account, benefiting from the available bandwidth in the broadband mmWave systems, the accurately estimated MPCs' TDoA information can be further utilized to refine the above location sensing result, whose implementation will be detailed next.    

Similar to the angle and distance estimation in \eqref{eq:power specturm orientation}, the relative delay of the $l$-th multipath is firstly obtained by frequency-domain matching filter as
\begin{align}\label{eq:frequency MF}
	\hat{\tau}_l  =  \frac{1}{N_{\rm UT}} \sum_{n=1}^{N_{\rm UT}} \mathop{\arg\max}\limits_{\tau[n]}
	{\left\Vert  \left\{  \widehat{\mathbf{H}}_{\rm UL}^{P} [m] \right\}_{f(\hat{\theta}_l^{\rm BS},\hat{r}_l^{\rm BS}),n,1:m}
		\, \mathbf{a}_{f}^* (\tau[n])  \right\Vert^2},   
\end{align}
where $\mathbf{a}_f (\tau) = e^{-j 2 \pi \tau [f_1,f_2,\dots,f_M]^{\rm T}} \in \mathbb{C}^{M \times 1}$ represents the frequency-domain channel response parameterized by relative channel delay $\tau$, while $f_1,f_2,\dots,f_M$ are the frequencies of the first, second and $M$ subcarriers, respectively.
Considering the different geographical locations of transceivers and the limited clock synchronization accuracy, it is difficult to measure the absolute delay of MPCs, and thus it is more effective to exploit the TDoA of MPCs 
\begin{align} \label{eq:TDoA}
	\hat{\tau}_l^{\rm TDoA} = \hat{\tau}_l - \frac{\sum_{l^{\prime} \in \mathcal{G}} \hat{\tau}_{l^{\prime}}}{|\mathcal{G}|_c} , l \in \mathcal{S}.
\end{align}
In addition, the TDoA of MPCs can also be calculated according to the location sensing results
\begin{align}
	 \bar{\tau}_l^{\rm TDoA}  = \frac{\sqrt{\hat{x}_l^2 + \hat{y}_l^2} + \sqrt{(\hat{x} - \hat{x}_l)^2 + (\hat{y} - \hat{y}_l)^2} - \sqrt{\hat{x}^2 + \hat{y}^2}}{c},
\end{align}
which is the function of coordinates $(\hat{x},\hat{y}),(\hat{x}_l,\hat{y}_l)$.
Owing to the location sensing errors, there exists difference between the calculated TDoA of MPCs and the real TDoA of MPCs, and thus define the loss function as
\begin{align}
	\mathcal{L}(  \hat{x}, \hat{y} ) = \sum_{l \in \mathcal{S}} \left| \hat{\tau}_l^{\rm TDoA} -  \bar{\tau}_l^{\rm TDoA} ( \hat{x}, \hat{y} ) \right|^2,
\end{align}

Following the maximum likelihood principle, the optimal UT and scatterers coordinates tend to minimize the above loss function, but it is intractable to directly derive its optimal closed-form solution. 
To this end, a feasible solution is to use gradient descent algorithm iteratively update UT and scatterers coordinates $(\hat {x},\hat{y}) $ and $(\hat{x}_l, \hat{y}_l)_{l \in \mathcal{S}}$\cite{lzr}.
Therefore, in the $t$-th iteration, one of the parameters in the coordinates is updated each time, while the other parameters are hold unchanged.
For instance, the update of the UT's coordinate can be represented as 
\begin{subequations}\label{eq:gradient descent}
	\begin{align}
		\hat{x}^{t} = \hat{x}^{t-1} -  \Delta_x \cdot \left. \frac{\partial \, \mathcal{L}(  \hat{x}, \hat{y}^{t-1}  )}{\partial \, \hat{x}} \right |_{\hat{x} = \hat{x}^{t-1}} , \\
		\hat{y}^{t} = \hat{y}^{t-1} -  \Delta_y \cdot \left. \frac{\partial \, \mathcal{L}(  \hat{x}^{t-1}, \hat{y}  )}{\partial \, \hat{y}} \right |_{\hat{y} = \hat{y}^{t-1}}.
	\end{align}
\end{subequations}
To guarantee the convergence of the loss function, an effective approach is Armijo-Goldstein backtracking line
search \cite{optimization} to determine the update step size $\Delta_x$ and $\Delta_y$.
Besides, the location of scatterers can be updated similar to \eqref{eq:gradient descent}.
In this case, with a small number of iterations, the refined coordinates are inclined to converge to the appoximately optimal solution. 
So far, we have finished the discussion about the proposed location sensing scheme for UM-MIMO systems, and its complete procedures are summarized in Alg. \ref{alg:localization}. 

{ In summary, our approach enhances localization accuracy by leveraging the TDOA technique with multiple scatterers as virtual anchors.  By calculating the time differences of arrival from these scatterers to the user, and incorporating the scatterers' known positions obtained through Alg. \ref{alg:localization}, our method refines the localization results. This supplementary use of TDOA, despite the presence of only a single BS, effectively enhances the robustness of the localization in environments with significant multipath effects.}

\begin{algorithm}[t] 	
	\caption{Proposed Location Sensing Scheme}
	\label{alg:localization}
	\LinesNumbered 
	\KwIn{Received reference signal ${\mathbf{Y}}_{\rm UL}[m]$, estimated sparse channel matrix $\widehat{\mathbf{H}}_{\rm UL}^{P}[m]$, the average noise variance $\overline{\hat{\sigma}^2} [m]$, and the maximum refinement iteration number $T_{\rm grd}$.}  
	
	\KwOut{The estimated UT and scatterers' position $(\bar{x},\bar{y})$ and $(\bar{x}_{l},\bar{y}_{l}), l \in \mathcal{S}$.}
	
	Initialize to estimate the number of MPCs $\hat{L}^{\prime}$ in accordance with  \eqref{eq:auto-correlation}-\eqref{eq:L estimate}.  
	
	\textbf{VAs Location Mapping:}
	
	Identify effective scatterers grids set $(\tilde{\varTheta},\tilde{\mathcal{R}})$ following \eqref{eq:eff grid}.	
	
	Utilize K-Means algorithm to cluster scatteres grids subset $(\tilde{\varTheta}_l,\tilde{\mathcal{R}}_l)$ for the $l \, (1 \le l \le \hat{L}^{\prime})$-th MPC.	
	
	Identify the truthful VAs coordinates $(\hat{\theta}_l^{\rm BS},\hat{r}_l^{\rm BS})$ in \eqref{eq:grid match}, and distinguish scatterers and subarray center set $\mathcal{S}$ and $\mathcal{G}$. 
	
	\textbf{Coarse UT Location Sensing:}
	
	Estimate AoDs $\hat{\theta}_l^{\rm UT}$ for identified VAs following 
	\eqref{eq:power specturm orientation}.
	
	Sense UT coarse location $(\hat{x}, \hat{y})$ exploiting VAs in \eqref{eq:wls}.   
	
	\textbf{Sensing Results Refinement:}
	
	Measure TDoA for identified MPCs through \eqref{eq:frequency MF}-\eqref{eq:TDoA}. 
	
	\For{$1 \le t \le T_{\rm grd}$}{

		Calculate the gradient and Armijo-Goldstein backtracking line
		search step size $\Delta_x$ and $\Delta_y$.
			
		Update the coordinates of UT and scatterers by \eqref{eq:gradient descent}.

	}
	
	\KwResult{The refined UT and scatterers' positions $({\bar{x}},{\bar{y}}) = (\bar{x}^{T_{\rm grd}},\bar{y}^{T_{\rm grd}})$ and $({\bar{x}}_l,{\bar{y}}_l) = (\bar{x}^{T_{\rm grd}}_{l},\hat{y}^{T_{\rm grd}}_{l}), l \in \mathcal{S}$.}
\end{algorithm}

\section{Hybrid Beamforming Design for Downlink Data Transmission}

To fully unleash the potential of UM-MIMO for more efficient data transmission, this section designs a beamforming scheme attempting to mitigate the spectral efficiency degradation caused by HFBS issue.
   
\subsection{Analog Codebook Design}

For the downlink data transmission stage, the received baseband data ${\mathbf{y}}_{\rm DL}[m] \in \mathbb{C}^{N_s^d}$ at the UT on the $m$-th subcarrier can be formulated as the following problem similar to \eqref{eq:UL-CE} as\footnote{It is assumed that the BS serves the different UTs using orthogonal time-frequency resources, and thus the beamforming design is optimized for a single specific UT to disclose the impact of HFBS issue.}    
\begin{align}\label{eq: DL}
 	\mathbf{y}_{\rm DL}[m] = \mathbf{W}^{\rm H}_{\rm DL}[m] \mathbf{H}_{\rm DL}[m] \mathbf{F}_{\rm DL}[m] \mathbf{s}[m] + \mathbf{W}^{\rm H}_{\rm DL}[m] \mathbf{n}[m],
\end{align}
Different from the parameters defined in \eqref{eq:UL-CE}, $\mathbf{F}_{\rm DL}[m]$ represents the BS's transmit beamforming matrix $\mathbf{F}_{\rm DL}[m] = \mathbf{F}^{\rm RF}_{\rm DL} \mathbf{F}^{\rm BB}_{\rm DL}[m] \in \mathbb{C}^{N_{\rm BS} \times N_s^d}$ with the power constraint $\Vert \mathbf{F}_{\rm DL}[m] \Vert_F^2 \le P_t^{\rm DL}/M$, in which $\mathbf{F}^{\rm RF}_{\rm DL} \in \mathbb{C}^{N_{\rm BS} \times N_{\rm BS}^{\rm RF}}$ and $\mathbf{F}^{\rm BB}_{\rm DL}[m] \in \mathbb{C}^{N_{\rm BS}^{\rm RF} \times N_s^d}$ are the analog and digital beamforming matrices, respectively, and $N_s^d$ is the number of simultaneous transmitted data streams. 
Similarly, $\mathbf{W}_{\rm DL}[m]$ denotes the UT's receive combining matrix $\mathbf{W}_{\rm DL}[m] = \mathbf{W}^{\rm RF}_{\rm DL} \mathbf{W}^{\rm BB}_{\rm DL}[m] \in \mathbb{C}^{N_{\rm UT} \times N_s^d}$, where $\mathbf{W}^{\rm RF}_{\rm DL} \in \mathbb{C}^{N_{\rm UT} \times N_{\rm UT}^{\rm RF}}$ and $\mathbf{W}^{\rm BB}_{\rm DL}[m] \in \mathbb{C}^{N_{\rm UT}^{\rm RF} \times N_s^d}$ are the analog and digital combining matrices, while $\mathbf{s}[m]$ is the payload data stream. 
In the data transmission stage, we seek to optimize the precoder and combiner that
maximizes the beamforming gain over the whole bandwidth and enhance the spectral efficiency (SE) expressed in \eqref{eq:SE} shown at the top of the next page, where $\mathbf{R}_n [m] = \sigma^2 \mathbf{W}^{\rm H}_{\rm DL}[m]  \mathbf{W}_{\rm DL}[m]$ is the noise covariance matrix after combining.  

\begin{figure*}[t]	
	\begin{align}\label{eq:SE}
	 R = \frac{1}{M} \sum_{m = 1}^{M} 
	 \log_2 \left( \left| \mathbf{I}_{N_s^d} +    \mathbf{R}_n^{-1}[m]  \mathbf{W}^{\rm H}_{\rm DL}[m] \mathbf{H}_{\rm DL}[m] \mathbf{F}_{\rm DL}[m]  \mathbf{F}_{\rm DL}^{\rm H}[m]  \mathbf{H}_{\rm DL}^{\rm H}[m] \mathbf{W}_{\rm DL}[m]  \right| \right)
	\end{align}
	\setcounter{equation}{46}
	\begin{align} \label{eq:47}
	g_0 (r,\theta,m) = \frac{1}{N_{\rm BS}} 
	\left|  \sum_{n=1}^{N_{\rm BS}} e^{j \left( \frac{2\pi}{\lambda_m} \frac{\delta_{n}^2 d^2 \cos^2 \theta}{2r} -  \frac{2\pi}{\lambda_c} \frac{\delta_{n}^2 d^2 \cos^2 \theta_0}{2r_0}  \right)
		- j \left(
		\frac{2\pi}{\lambda_m} \delta_{n} d \sin \theta -
		\frac{2\pi}{\lambda_c} \delta_{n} d \sin \theta_0
		\right)
	}  \right|
	\end{align}
	\hrulefill
\end{figure*}

Overcoming beam squint requires that each RF phase shifter generate different phase shift across different subcarriers to form frequency-dispered beamforming vectors.
However, the RF phase shifter can only generate the frequency-independent phase shift, greatly limiting its beamforming ability especially in the presence of beam-squint effect.
In this case, traditional analog narrow beams design approach that focus on optimizing for center-frequency suffers from serious performance degradation. 
Therefore, elaborate design of the analog beams codebook is pivotal to an efficient beamforming scheme for hybrid UM-MIMO systems in the presence of HFBS effect.

Without loss of generality, we assume that the reference frequency for designing the analog phase shifts equals the center frequency $f_c$.
Particularly, in the case of $N_{\rm UT} = N_{\rm BS}^{\rm RF} = 1$, when the  beamforming $\mathbf{f}_{\rm DL}^{\rm RF}(r_0,\theta_0)$ is designed to focus on the postion $(r_0,\theta_0)$ for the reference frequency, then $\mathbf{f}_{\rm DL}^{\rm RF}(r_0,\theta_0) = \frac{1}{\sqrt{N_{\rm BS}}} e^{-j \frac{2\pi} {\lambda_c} [r_{0,1},r_{0,2},\dots,r_{0,N_{\rm BS}}]^{\rm T}}$, where $r_{0,n} = \sqrt{ r_0^2 + \delta^2_{n} d^2 - 2 r_0 \delta_{n} d \sin\theta_0 }$.
In this case, the beamforming gain at the given position $(r,\theta)$ on the $m$-th subcarrier can be computed as  
\setcounter{equation}{45}
\begin{align} \label{eq:beamforming gain}
	g_0 (r,\theta,m)  & = \left| \mathbf{a}_{\rm BS}^{\rm H}(r,\theta, m) \mathbf{f}_{\rm DL}^{\rm RF}(r_0,\theta_0) \right|, \notag \\
	& = \frac{1}{N_{\rm BS}} \left| \sum_{n=1}^{N_{\rm BS}} e^{j \frac{2\pi}{\lambda_m} r_{n} -  j \frac{2\pi}{\lambda_c} r_{0,n}} \right|.
\end{align} 
By means of Fresnel approximation mentioned in \eqref{eq:Fresnel approximation},  \eqref{eq:beamforming gain} can be further derived as \eqref{eq:47}, which is shown at the top of the next page. 
\begin{figure*}[t]	
	\centering
	\subfigure[Trajectory of beam-squint positions across different subcarriers.]{	
		\includegraphics[width=0.6\columnwidth, keepaspectratio]{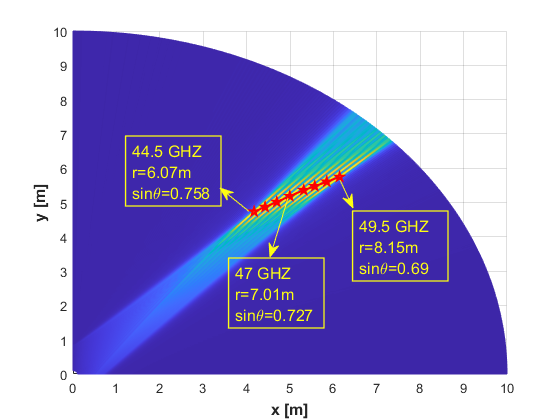}
		\label{fig5:trajectory}
	}	
	\subfigure[Beamforming gain variation versus angles at the optimal distance $g_0 (\tilde{r}_m,\theta,m)$.]{
		\includegraphics[width=0.6\columnwidth, keepaspectratio]{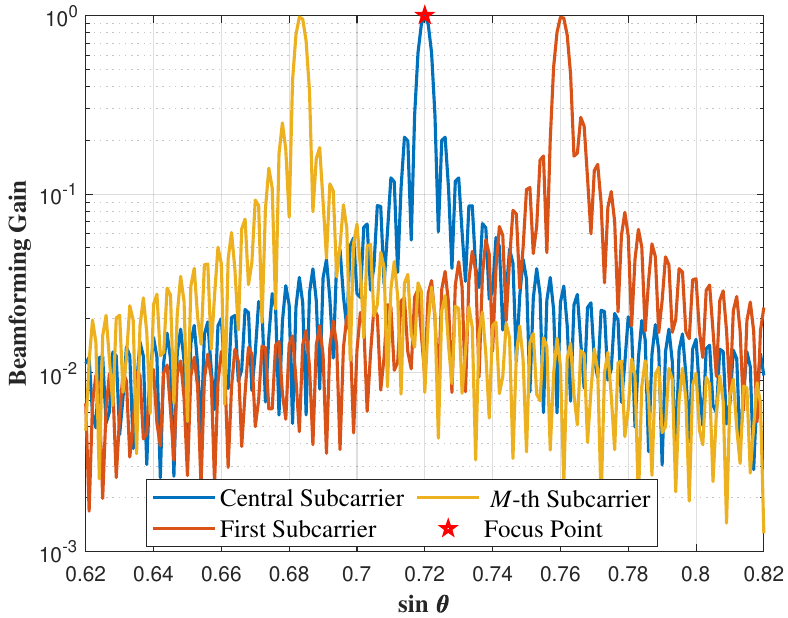}
		\label{fig5:distance}
	}
	\subfigure[Beamforming gain variation versus distances at the optimal angle $g_0 (r,\tilde{\theta}_m,m)$.]{	
		\includegraphics[width=0.6\columnwidth, keepaspectratio]{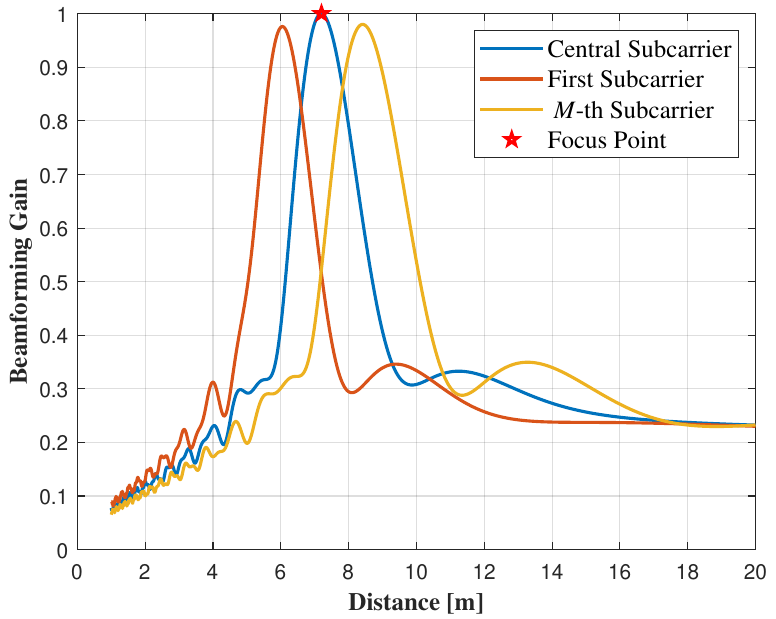}
		\label{fig5:angle}
	}	
	\captionsetup{font={color = {black}}, justification = raggedright,labelsep=period}
	\caption{Illustration of hybrid-field beam-squint effect, where the system parameters are $f_c = 47\rm GHz$, $\text{BW}=5\text{GHz}$, $N_{\rm BS} = 512$, $N_{\rm UT}=N_{\rm BS}^{\rm RF} = 1$, and $M = 8$ without regard to large-scale path loss for easy demonstration. The transmit beamforming is set to focus on the location $(r,\theta) = (7.01m,46.6^{\circ})$ at the center frequency.}	
	\vspace{-5mm}	
\end{figure*}
Denote the beam-squint position that maximize the beamforming gain on the $m$-th subcarrier as $(\tilde{r}_m,\tilde{\theta}_m)$, which can be expressed as \cite{gff2}  
\setcounter{equation}{47}
\begin{subequations}\label{eq:beam-squint trajectory}
	\begin{align}
		 \sin \tilde{\theta}_m  & =  \sin \theta_0 \cdot \frac{\lambda_m}{\lambda_c}, \\
		 \tilde{r}_m & = r_0 \cdot \frac{\lambda_c}{\lambda_m} \cdot \frac{1- \sin^2 \tilde{\theta}_m}{1-\sin^2 \theta_0}.
	\end{align}
\end{subequations}
\eqref{eq:beam-squint trajectory} indicates that distinct from the far-field propagation, HFBS effct includes angle squint effect and distance squint effect, respectively.
Fig. \ref{fig5:trajectory} illustrates the beamforming gain distribution $g_0 (r,\theta,m)$ with different positions and subcarriers $(r,\theta,m)$ via a heatmap, and the beam-squint trajectory of $(\tilde{r}_m,\tilde{\theta}_m)$ plotted by a red-star line. 
To further provide insights into analog beamforming design, we investigate to decouple the variation of beamforming gain along distances or angles, i.e., $g_0 (\tilde{r}_m,\theta,m)$ and  $g_0 (r,\tilde{\theta}_m,m)$, respectively, and 
disclose the sepreate effects of angle squint and distance squint.
Substitute the beam-squint positions derived in \eqref{eq:beam-squint trajectory}, 
 $g_0 (\tilde{r}_m,\theta,m)$ and  $g_0 (r,\tilde{\theta}_m,m)$ can be expressed as \cite{cmy}
\begin{subequations}
	\begin{align}	 
		g_0 (\tilde{r}_m,\theta,m) & = \left|  \frac{1}{N_{\rm BS}}  \sum_{n=1}^{N_{\rm BS}} e^{
			- j \left(
			\frac{2\pi}{\lambda_m} \delta_{n} d \sin \theta -
			\frac{2\pi}{\lambda_c} \delta_{n} d \sin \theta_0
			\right)
		}  \right| \notag \\	
		& =
		\left|  \frac{\sin \left(\frac{\pi d N_{\rm BS}}{\lambda_m}\sin\theta  - \frac{\pi d N_{\rm BS}}{\lambda_c}\sin\theta_0 \right)}{N_{\rm BS} \sin \left(\frac{\pi d}{\lambda_m}\sin\theta  - \frac{\pi d}{\lambda_c}\sin\theta_0 \right)}    \right|, 
 	\end{align}
	\begin{align}
		g_0 (r,\tilde{\theta}_m,m) & = \left|  \frac{1}{N_{\rm BS}} \sum_{n=1}^{N_{\rm BS}} e^{j \left( \frac{2\pi}{\lambda_m} \frac{\delta_{n}^2 d^2 \cos^2 \tilde{\theta}_m}{2r} -  \frac{2\pi}{\lambda_c} \frac{\delta_{n}^2 d^2 \cos^2 \theta_0}{2r_0}  \right)
		}  \right| \notag \\
		\quad & \approx \left| \frac{1}{N_{\rm BS}} \int_{-N_{\rm BS}/2}^{N_{\rm BS}/2} e^{j \pi n^2 \left(  \frac{d^2 \cos^2 \tilde{\theta}_m }{\lambda_m r} -  \frac{d^2 \cos^2 \theta_0 }{\lambda_c r_0} \, 
		\right) } {\rm d} n \right|,
	\end{align}
\end{subequations}
which are displayed in Fig. \ref{fig5:distance} and  Fig. \ref{fig5:angle}, respectively.
On the one hand, it can be observed that the angle squint effect causes the beamforming gain of non-central subcarriers w.r.t. position $(r_0,\theta_0)$ to fall out of the main lobe and attenuate substantially.  
On the other hand, the distance squint effect also leads the received energy associated with non-central subcarriers on position $(r_0,\theta_0)$ to decline by half. 

Besides, according to the conclusion derived from \eqref{eq:beam-squint trajectory}, to enable the generated beam to focus on the expected location $(r_0,\theta_0)$ for the $m$-th subcarrier, $\mathbf{f}_{\rm DL}^{\rm RF}(r_m^{\star},\theta_m^{\star})$ should meet  
\begin{subequations}\label{eq:pre-squint}
	\begin{align}
		 \sin \theta_m^{\star}  & =  \sin \theta_0 \cdot \frac{\lambda_c}{\lambda_m}, \\
		 \frac{1}{r_m^{\star}} & = \frac{1}{r_0} \cdot \frac{\lambda_c}{\lambda_m} \cdot \frac{1- \sin^2 \theta_m^{\star}}{1-\sin^2 \theta_0}.
	\end{align}
\end{subequations}
Therefore, to provide constant beamforming gain as much as possible over the whole bandwidth and maximize SE, one intuitive compromise is to generate beams $\mathbf{b}(r_0,\theta_0)$ covering all $(r_m^{\star},\theta_m^{\star}), 1\le m \le M$ so that the traditional narrow beams are required to be broadened along both the distance and angle at the reference frequency, with the maximum angle and distance intervals $\Delta \sin \theta_0 = |\sin \theta_M^{\star} - \sin \theta_1^{\star}|$ and $\Delta \frac{1}{r_0} = | \frac{1}{r_M^{\star}} - \frac{1}{r_1^{\star}} |$, respectively. 
	
The generation of beam $\mathbf{b}(r_0,\theta_0)$ can employ a subarray based beam aggregation method \cite{gff}, which divides the antenna array into $G = \lceil \frac{N_{\rm BS}}{N_{\rm sub}} \rceil$ subarrays and each with $N_{\rm sub}$ elements, and synthesizes the sub-beams generated by different subarrays.
Specifically, according to the array signal processing theory, the 3dB beam width of an $N_{\rm sub}$-elements subarray is 
$\Delta_{\rm sub} \sin \theta  = \frac{1}{N_{\rm sub}}$  and
$\Delta_{\rm sub} \frac{1}{r} = \frac{1.556}{N_{\rm sub}^2 \lambda_c \cos^2 \theta_0}$	along the normalized angle and distance domains, respectively \cite{cmy-mag}.
In line with the requirements for beam broadening region derived in  \eqref{eq:pre-squint}, the number of subarrays $G$ should satisfy
\begin{subequations}
	\begin{align}
	 & \frac{1}{N_{\rm sub}} \cdot G 	 \ge \Delta \sin \theta_0 , \\
	 & \frac{1.556}{N_{\rm sub}^2 \lambda_c \cos^2 \theta_0} \cdot G  \ge \Delta \frac{1}{r_0}.
\end{align}                                                                                \end{subequations}   
And the beamforming vector $\mathbf{b}(r_0,\theta_0)$ can be generated subarray-wisely for the $g$-th virtual subarray as 

\begin{align}
	&{\left[ \mathbf{b}(r_0,\theta_0) \right]_{(g-1)N_{\rm sub} + n} = \notag} \\ 
	& {\frac{1}{\sqrt{N_{\rm BS}}} e^{-j \frac{2\pi}{\lambda_c} \left[ -\delta_{[(g-1)N_{\rm sub} + n]} \sin \theta_g + \frac{\delta_{[(g-1)N_{\rm sub} + n]}^2 d^2 \cos^2 \theta_v}{2 r_g} \right] } ,}
\end{align}
where $\sin \theta_g = \sin \theta_0 + \frac{2g-1}{2N_{\rm sub}}$ and $\frac{1}{r_g} = \frac{1}{r_0} + \frac{0.778(2g-1)}{N_{\rm sub}^2 \lambda_c \cos^2 \theta_0}$. 
In this problem, in order to maximize the beam gain and improve SE, the beam energy should be focused on UT or the scatterers as much as possible.
Therefore, the analog beamforming codebook used to optimize RF phase shifts can be constructed as  
$\mathbf{B} = \left\{ \mathbf{b} (\bar{r}^{\rm BS},\bar{\theta}^{\rm BS}) \right\} \cup \left\{ \mathbf{b} (\bar{r}_l^{\rm BS},\bar{\theta}_l^{\rm BS}) | \, l \in \mathcal{S} \right\} \cup \left\{ \mathbf{b} (\hat{r}_l^{\rm BS},\hat{\theta}_l^{\rm BS}) | \, l \in \mathcal{G} \right\} \in \mathbb{C}^{N_{\rm BS} \times (\hat{L}^{\prime}+1)}$,
which collects all UT's estimated location, scatterers' location, and the estimated UT's subarray center, while $\bar{r}^{\rm BS} = \sqrt{\bar{x}^2 + \bar{y}^2}$, $\bar{\theta}^{\rm BS} = \text{arctan}(\bar{y}/\bar{x})$ and $\bar{r}_l^{\rm BS} = \sqrt{\bar{x}_l^2 + \bar{y}_l^2}$, $\bar{\theta}_l^{\rm BS} = \text{arctan}(\bar{y}_l/\bar{x}_l)$, respectively.  
        
\subsection{Hybrid Beamforming Design with Designed Codebook}

Thanks to the mmWave UM-MIMO channel sparsity, the designed analog beamforming codebook is capable of approximately spaning the column space of the unconstrained full-digital beamforming matrix $\mathbf{F}_{\rm DL}^{\rm opt} [m]$ on the m-th subcarrier for $1 \le m \le M$, where $\mathbf{F}_{\rm DL}^{\rm opt} [m]$ composes of the first $N_s^d$ columns of the right singular matrix of $\mathbf{H}_{\rm DL}[m]$ \cite{omp-precoding}. 
Therefore, the hybrid beamforming design problem under the HFBS effect is equivalent to selecting the $N_{\rm BS}^{\rm RF}$ best-matched RF beamforming vectors and finding their optimal baseband combination, which can be concretely expressed as 
\begin{align}
		 \left( \tilde{\boldsymbol{\Lambda}}, \tilde{\mathbf{F}}_{\rm DL}^{\rm BB} [m]  \right) &  =
		 \mathop{\arg\min}_{\boldsymbol{\Lambda}, \mathbf{F}_{\rm DL}^{\rm BB} [m]} \,
		\sum_{m=1}^{M}
		\Vert
		\mathbf{F}_{\rm DL}^{\rm opt} [m] - \mathbf{B} \boldsymbol{\Lambda} \mathbf{F}_{\rm DL}^{\rm BB} [m]
		\Vert_{F},  \notag  \\ 
		&   {\rm s.t.} \quad \Vert {\rm diag} \left( \boldsymbol{\Lambda} \boldsymbol{\Lambda}^{\rm H} \right) \Vert_0 = N_{\rm BS}^{\rm RF}, \notag \\
		&  \quad\quad\; \Vert \mathbf{B} \boldsymbol{\Lambda} \mathbf{F}_{\rm DL}^{\rm BB} [m] \Vert_F^2 \le P_t^{\rm DL}/M,
\end{align}
where $\boldsymbol{\Lambda}$ is a selection matrix used to select beamforming vectors from the codebook $\mathbf{B}$ for constructing the analog beamformer as $\mathbf{F}_{\rm DL}^{\rm RF} = \mathbf{B} \boldsymbol{\Lambda}$.
Considering the sparsity of selection matrix $\boldsymbol{\Lambda}$, the resulting 
problem formulation is identical to the optimization problem of sparse signal recovery. 
Therefore, we straightforwardly exploit the method based on the classic concept of SOMP \cite{omp-precoding} to obatain the solution of $\left( \tilde{\mathbf{F}_{\rm DL}^{\rm RF}},  \tilde{\mathbf{F}}_{\rm DL}^{\rm BB} [m] \right), 1 \le m \le M$. 
For the receiver at the UT, the optimal full-digital combiner at each subcarrier is the MMSE combiner $\mathbf{W}_{\rm DL}^{\rm MMSE} [m]$ in the absence of any hardware restrictions \cite{omp-precoding}.  
And the hybrid combiner design  $\left( \tilde{\mathbf{W}_{\rm DL}^{\rm RF}},  \tilde{\mathbf{W}}_{\rm DL}^{\rm BB} [m]  \right) , 1 \le m \le M$ at the receiver can be acquired in a similar way to approximate $\mathbf{W}_{\rm DL}^{\rm MMSE} [m]$.

\section{Simulation Experiments}

This section conducts extensive simulation experiments to validate the superiority of our proposed ILSC schemes and compares theri performance with the counterparts in previous literature.

\subsection{Simulation Setup}

First of all, we present the main system parameters used in simulation experiments in TABLE \ref{tab:my-table} \footnote{{ Tens up to a hundred GHz will be obtained in THz band for 6G and beyond era\cite{2023_IEEENetwork_NanYang_ChongHan}.
            Although the current 5G NR does not have continuous 5 GHz bandwidth, this parameter setting is used to verify the effectiveness of our proposed ILSC scheme, which can be applied to address the HFBS problem in future networks.}} unless stated otherwise.
In addition, the noise power spectrum density at the receivers is $N_0 = -174 \, \rm{dBm/Hz}$, and the power of AWGN is accordingly calculated as $\sigma^2 = N_0 \times \rm{BW} = -77 \, \rm{dBm}$.
The UT's and scatterers' position can be simulated in line with the angle and distance distribution given in TABLE \ref{tab:my-table}.
For the MPCs resulted from scatterers, we employ the hybrid-field channel model in  \eqref{eq:channel model} to generate them.
Here, we only count single-hop paths and assume that the small-scale channel gain $\sigma_{\alpha}^2 = 1$. 
The large-scale path loss coefficients of MPCs are determined according to the empirical model provided in Table 7.4.1-1 of \cite{3gpp}. 
In the uplink, the transmit power of UT $P_t^{\rm UL}$ ensures the received signal-to-noise-ratio (SNR) to meet $30 \, \rm dB$.
{It is important to note that our study focuses on indoor and short-range localization and communication scenarios, rather than UMa environments. Thus, the simulation parameters are chosen to reflect these indoor settings.}

\begin{table}[t]
	\centering
	\caption{}
	\label{tab:my-table}
	\resizebox{0.6\columnwidth}{!}{%
		\begin{tabular}{@{}cc@{}}
			\toprule
			\textbf{Parameter}                                            & \textbf{Value}                              \\ \midrule
			Number of BS antenna $N_{\rm BS}$                             & 512                                         \\
			Number of UT antenna $N_{\rm UT}$                             & 32                                          \\
			Number of BS RF chains $N_{\rm BS}^{\rm RF}$                  & 4                                           \\
			Number of UT RF chains $N_{\rm UT}^{\rm RF}$                  & 4                                           \\
			Carrier frequency $f_c$                                       & $47 \, \rm GHz$                             \\
			System bandwidth ${\rm BW}$                                   & $5 \, \rm GHz$                             \\
			Number of subcarriers $M$                                     & 64                                       \\
			Maximum mumber of MPCs $L$                                            & 6                                           \\
			Scatterer angle $\theta_l^{\rm BS}$ distribution   & $\mathcal{U}[-\frac{\pi}{3},\frac{\pi}{3}]$ \\
			Scatterer distance $r_l^{\rm BS}$ distribution    & $\mathcal{U}[5,20](m)$                      \\
			{UT angle $\theta$ distribution }      & $\mathcal{U}[-\frac{\pi}{3},\frac{\pi}{3}]$ \\
			{UT distance $r$ distribution  }       & $\mathcal{U}[5,50](m)$                      \\
			Sampling lattice parameter $(\rho,S,\eta)$                    & $(2,20,1.5)$                                \\
			AMP damping factor $\epsilon$                                 & 0.8                                         \\
			AMP maximum iteration number $T_{\rm iter}$                   & 100                      \\
			Sensing refinement iteration number $T_{\rm iter}^{\prime}$                   & 10          	
			                   \\ \bottomrule
		\end{tabular}%
	}
\end{table}

\subsection{Channel Estimation Results}

Next, we investigate the performance of the proposed uplink CSI acquistion scheme by
evaluating the normalized mean square error (NMSE) between the real channel matrix $\mathbf{H}_{\rm UL}[m]$ and its estimation $\widehat{\mathbf{H}}_{\rm UL}[m]$, i.e., 
\begin{align}
	\text{NMSE} = \mathbb{E} \left\{ \frac{1}{M} \sum_{m=1}^{M}\frac{\Vert \mathbf{H}_{\rm UL}[m] - \widehat{\mathbf{H}}_{\rm UL}[m] \Vert_F^2}{\Vert  \mathbf{H}_{\rm UL}[m] \Vert_F^2} \right\}.
\end{align}
For comparison, we consider the baseline schemes including
\begin{enumerate}
	\item[(i)] traditional far-field DFT projection matrix based schemes using OMP algorithm \cite{omp-ce};
	\item[(ii)] far-field frequency-selective projection matrix based schemes with beam-squint effect into consideration \cite{beam-squint CE};
	\item[(iii)] subarray piecewise far-field approximate projection matrix based schemes \cite{myk};
	\item[(iv)] near-field polar-domain projection matrix based schemes using OMP algorithm with the oracle information of MPCs number $(L+1)$ \cite{cmy} or AMP algorithm.
\end{enumerate}

\begin{figure}[t]	
	\centering
	\subfigure[The performance comparison of different channel
	estimation schemes against different pilot length]{
		\includegraphics[width=0.46\columnwidth, keepaspectratio]{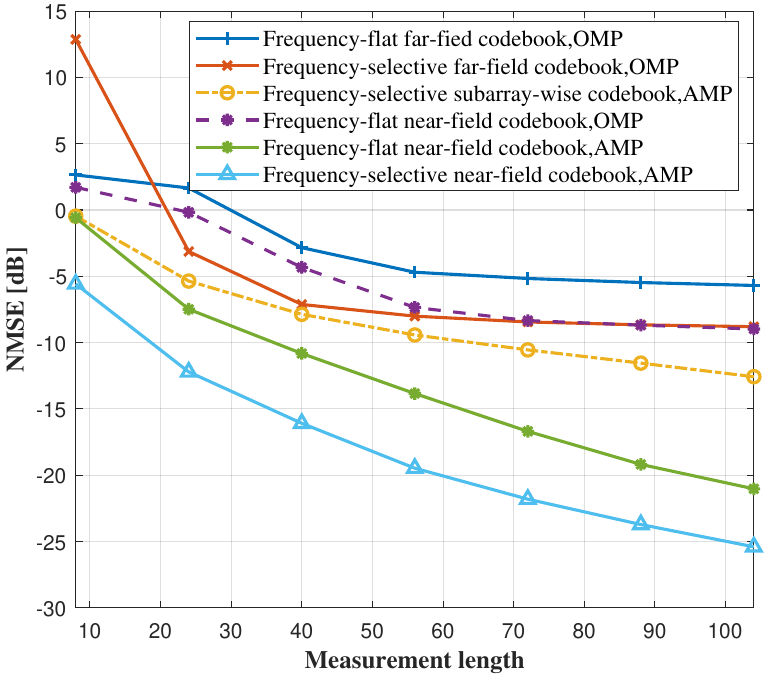}
		\label{fig:CE_pilot_length}
	}
	\subfigure[The performance comparison of different channel
	estimation schemes against different scatterers distance]{
		\includegraphics[width=0.46\columnwidth, keepaspectratio]{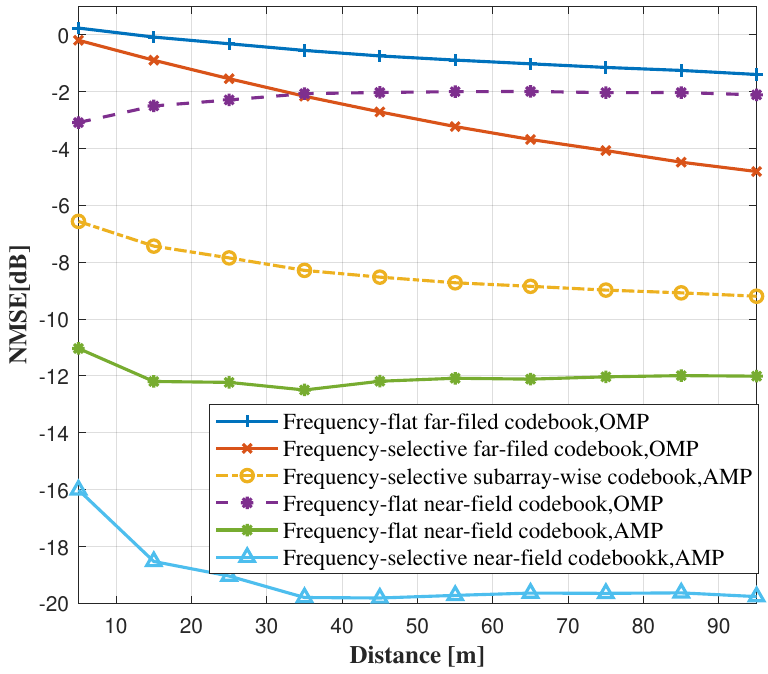}
		\label{fig:CE_distance}
	}	
	\captionsetup{font={color = {black}}, justification = raggedright,labelsep=period}
	\caption{Channel estimation performance comparison.}	
	\label{fig6:CE results}
	\vspace{-2mm}	
\end{figure}

Fig.~\ref{fig6:CE results} presents channel estimation performance comparison results versus measurement length $P$ and scatterers distance respectively, while in Fig. \ref{fig:CE_distance}, the measurement length is $P=72$.
On the one hand, the performance degradation of traditional far-field sparse channel estimation scheme proves that the channel characteristics in hybrid-field region is different from those in conventional far-field region, and the adopted hybrid-field projection matrix derived based on the Fresnel approximation can better capture its sparsity especially when the scatterers' or UT's distance is small. 
Although dividing the UM-MIMO array into multiple small subarrays and approximate their channel with far-field planar wave is a feasible solution, we still observe that its performance enhancement is limited as it increases the number of angle parameters that need to be estimated to some extent. 
On the other hand, the significant performance loss caused by the traditional frequency-flat projection matrices illustrates that the beam-squint effect destroys the common support set property of the channels, while utilizing the proposed frequency-dependent projection can address this issue. 
Moreover, the adopted AMP-EM algorithm shows evident superiority over OMP algorithm since it better exploits the statistical prior information of HFBS channel and AWGN, and  can also learn the information of MPCs' number adaptively. 

\begin{table}[t]

\renewcommand{\arraystretch}{1.5}
\centering
\small  
\resizebox{0.9\columnwidth}{!}{
\begin{tabular}{|c|c|c|c|c|}
\hline
\multirow{2}{*}{\diagbox{Method}{Metrics}}& \multicolumn{2}{c|}{Figure 7(a)} & \multicolumn{2}{c|}{Figure 7(b)} \\
\cline{2-5}
& $\textrm{RMSE}_\theta$[dB] &$\textrm{RMSE}_r$[dB] & $\textrm{RMSE}_\theta$[dB] & $\textrm{RMSE}_r$[dB] \\
\hline
Scatterers localization & {-11.2081} & {-5.6610} & {-10.4287} & {2.3688} \\
\hline
LoS-only localization & {-9.6738} & {4.1412} & {-5.4379} &{6.5823} \\
\hline
VAs-aided localization & {-9.3817} & {-1.8237} & {-16.5758} & {4.8653} \\
\hline
Refined Localization & {-9.6900} & {-16.9680} & {-17.6700} & {-22.4413} \\
\hline
\end{tabular}
}
\captionsetup{labelfont={color=black}, textfont={color=black}}
\caption{Comparison of RMSE values for different scatterers' localization errors.}
\label{tab:rmse_comparison}
\end{table}

\begin{figure}[t]	
	\centering
	\subfigure[Case that UT and scatterers are closer to the BS]{
		\includegraphics[width=0.46\columnwidth, keepaspectratio]{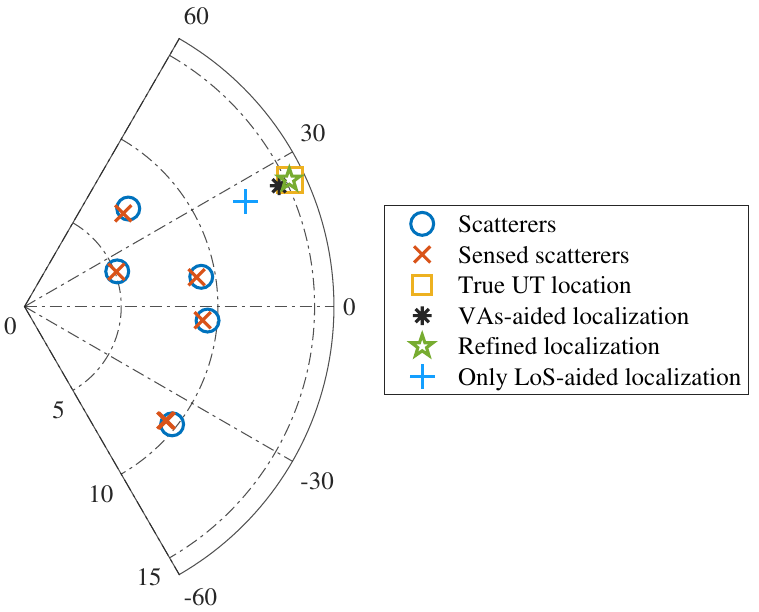}
		\label{fig:location_near}
	}
	\subfigure[Case that UT and scatterers are far away from the BS{, where one scatterer has been wrongly clustered}]{
		\includegraphics[width=0.455\columnwidth, keepaspectratio]{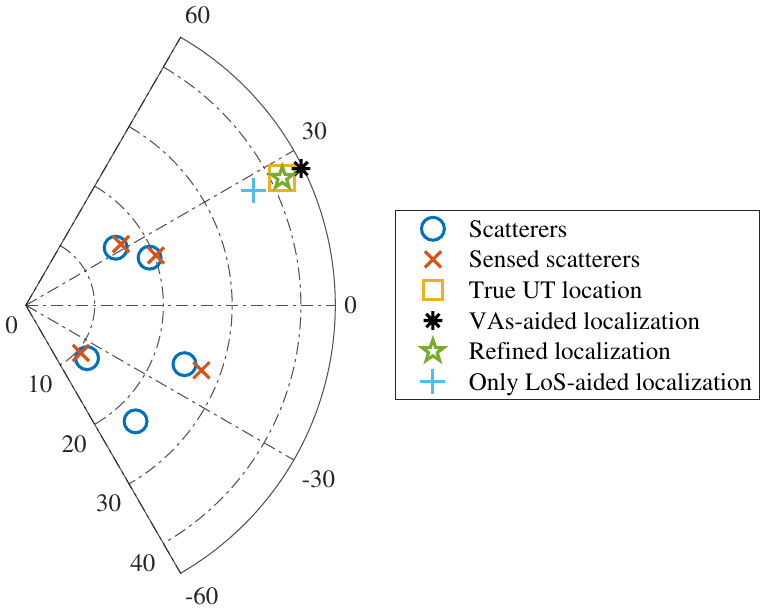}
		\label{fig:location_far}
	}	
	\captionsetup{font={color = {black}}, justification = raggedright,labelsep=period}
	\caption{A visualization of UT and scatterers location sensing performance.}	
	\label{fig7:location}
	\vspace{-3mm}	
\end{figure}

\begin{figure}[t]
	\vspace{-3mm}	
	\centering
	\subfigure[RMSE performance of UT's angle]{
		\includegraphics[width=0.47\columnwidth, keepaspectratio]{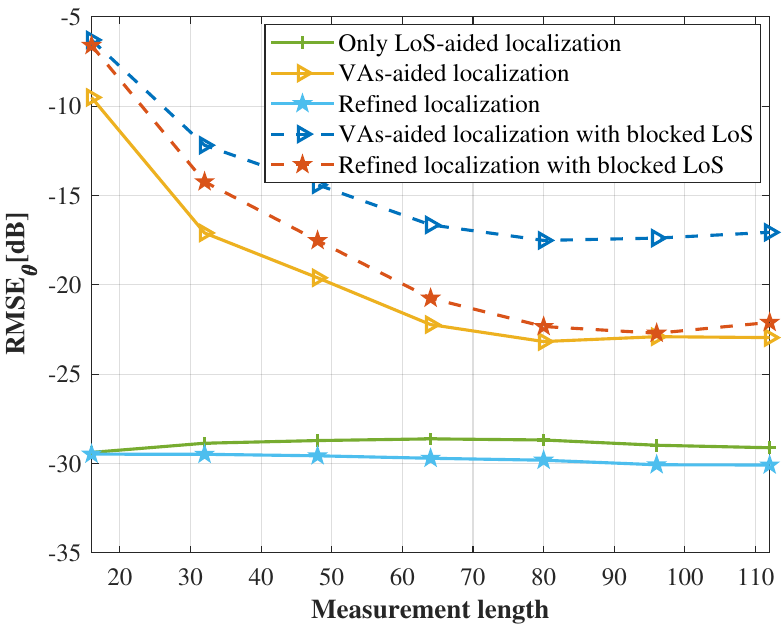}
		
	}	
	\subfigure[RMSE performance of UT's distance]{
		\includegraphics[width=0.46\columnwidth, keepaspectratio]{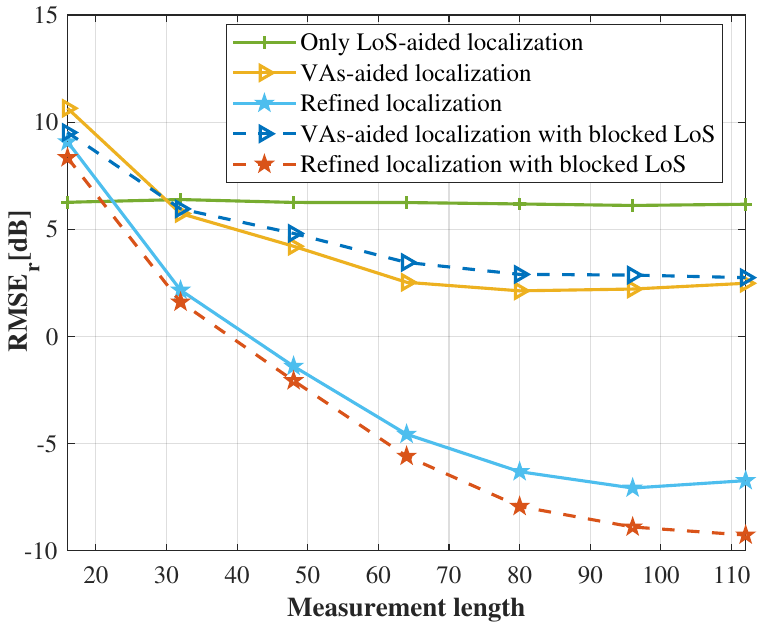}
	}
	\captionsetup{font={color = {black}}, justification = raggedright,labelsep=period}
	\caption{Location sensing performance against different pilot length.}
	\label{fig7:localization RMSE}	
\end{figure}

\subsection{Location Sensing Results}

\begin{figure}[t]
	\vspace{-5mm}	
	\centering
	\subfigure[CDF of ${\rm RMSE}_{\theta}$]{
		\includegraphics[width=0.46\columnwidth, keepaspectratio]{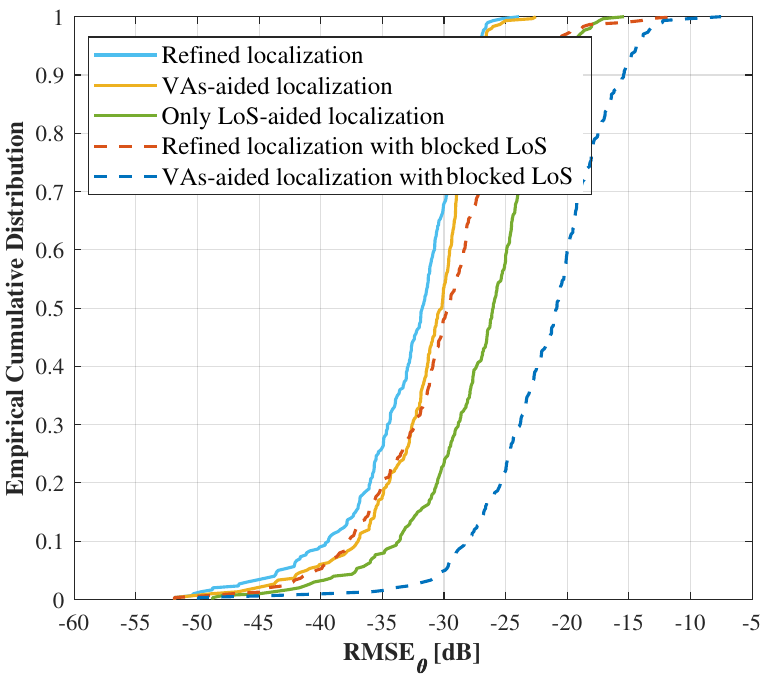}
	}	
	\subfigure[CDF of ${\rm RMSE}_r$]{		
		\includegraphics[width=0.46\columnwidth, keepaspectratio]{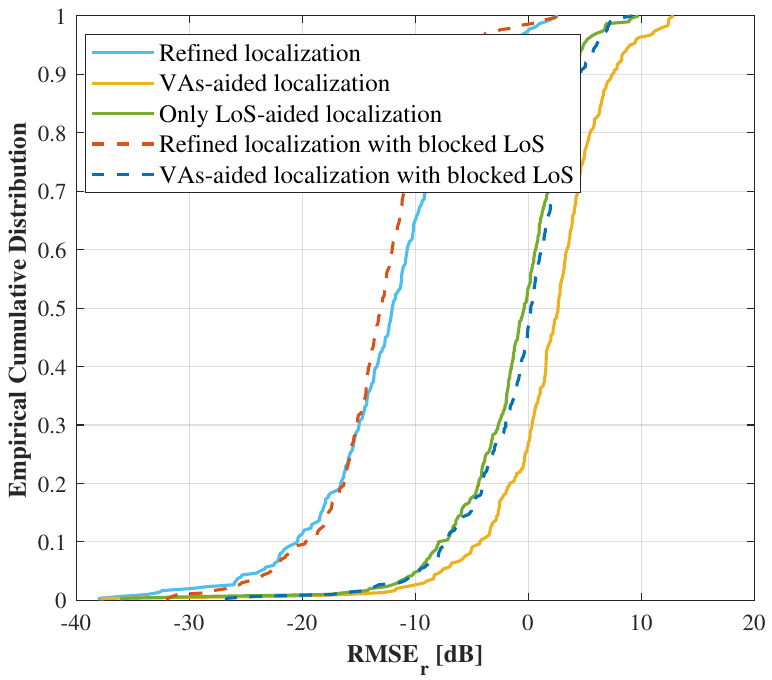}
	}
	\captionsetup{font={color = {black}}, justification = raggedright,labelsep=period}
	\caption{Empirical CDF w.r.t. the RMSE of angle and distance estimation.}	
	\label{fig:localization RMSE CDF}
	\vspace{-3mm}	
\end{figure}

\begin{figure}[t]
	\vspace{-3mm}	
	\centering
	\subfigure[{RMSE performance of UT’s angle}]{
		\includegraphics[width=0.46\columnwidth, keepaspectratio]{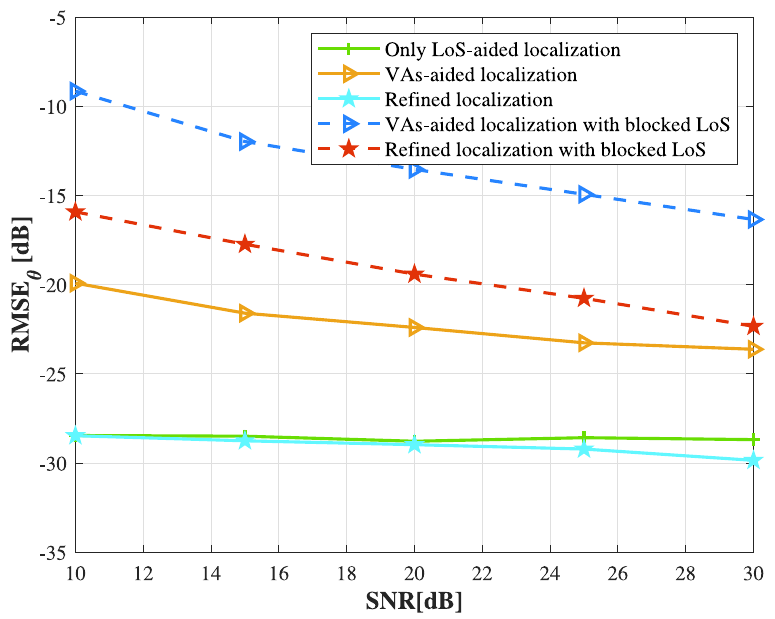}
	}	
	\subfigure[{RMSE performance of UT’s distance}]{		
		\includegraphics[width=0.46\columnwidth, keepaspectratio]{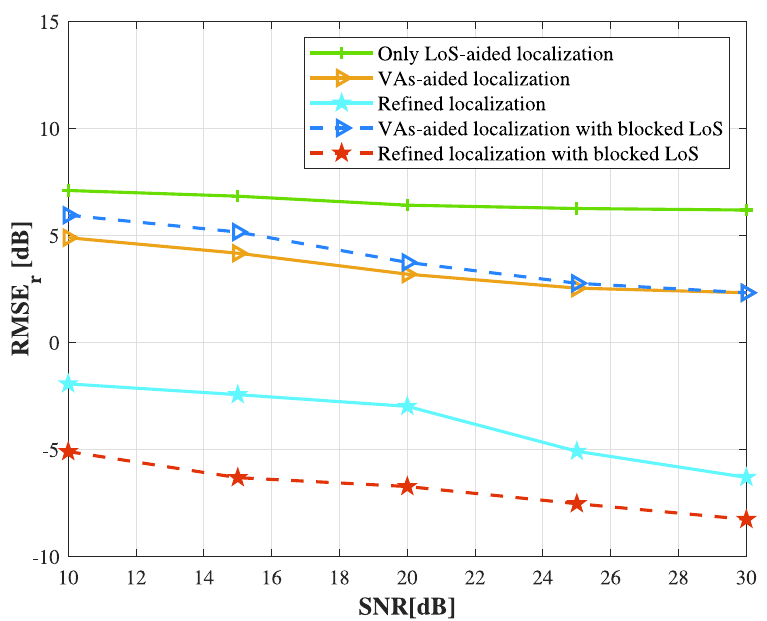}
	}
	\captionsetup{font={color = {black}}, justification = raggedright,labelsep=period}
	\caption{ Localization performance comparison against different uplink SNRs.}	
	\label{fig:localization RMSE SNR}
	\vspace{-5mm}	
\end{figure}

\setcounter{subfigure}{0}
\begin{figure}[htpb]

	\centering
	\subfigure[{RMSE performance of UT’s angle}]{
		\includegraphics[width=0.46\columnwidth, keepaspectratio]{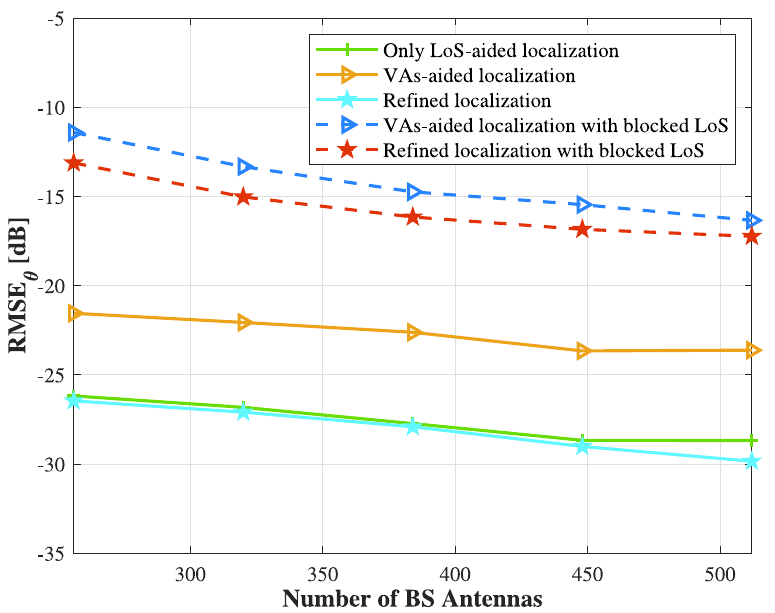}
	}	
	\subfigure[{RMSE performance of UT’s distance}]{		
		\includegraphics[width=0.45\columnwidth, keepaspectratio]{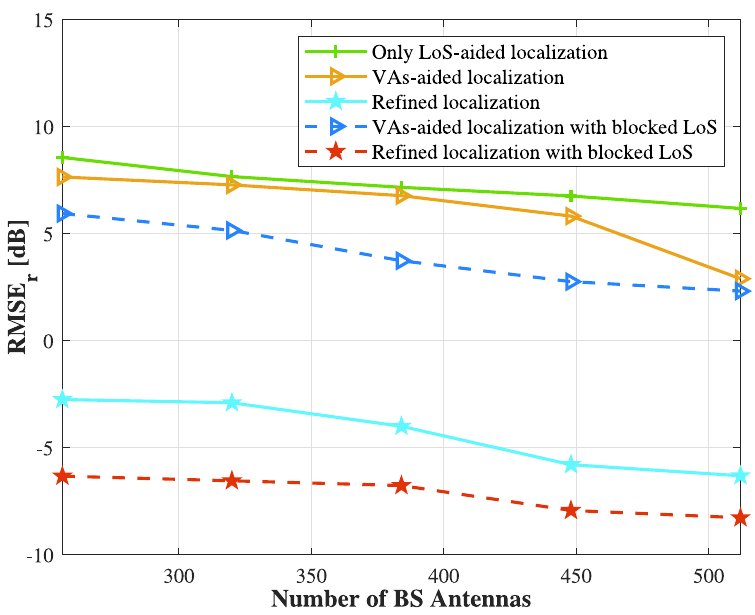}
	}
	\captionsetup{font={color = {black}}, justification = raggedright,labelsep=period}
 \captionsetup{font={color = {black}}, justification = raggedright,labelsep=period}
	\caption{\textcolor{black}{Localization performance comparison against different numbers of BS antennas.}}	
	\label{fig:localization RMSE BS}	
\end{figure} 

To evaluate the performance of location sensing, we adopt the root mean square error (RMSE) of UT's angle and distance estimation as the criteria, which can be defined as
\begin{subequations}
	\begin{align}
		& \text{RMSE}_{\theta} = \sqrt{\mathbb{E} \left\{ \left( \tilde{\theta}^{\rm BS} - \theta^{\rm BS} \right)^2  \right\} },  \\
		& \text{RMSE}_{r} = \sqrt{\mathbb{E} \left\{ \left( \tilde{r}^{\rm BS} - r^{\rm BS} \right)^2  \right\} },
	\end{align} 
\end{subequations}
In addition, we consider the following comparison schemes: 
\begin{enumerate}
	\item [(i)] location sensing with only LoS information \cite{Near-Field Localization};
	\item [(ii)] location sensing with the aid of VAs according to \eqref{eq:wls};
	\item [(iii)] Refined location sensing with TDoA of MPCs according to \eqref{eq:gradient descent}. 	
\end{enumerate}

In order to intuitively show the performance of location sensing, we compare the estimated locations of scatterers and UT with their true positions for a single realization with measurement length $P=72$ in Fig.~\ref{fig7:location}.
{ Besides, Table \ref{tab:rmse_comparison} shows the RMSE performance of the localization for scatterers and UT in Fig.~\ref{fig7:location}, and Table \ref{tab:rmse_comparison} could also demonstrate the impact of scatterers localization errors on various localization methods.}
In Fig.~\ref{fig:location_near} { and Table \ref{tab:rmse_comparison}}, we observe that when UT and scatterers are closer to the BS, all UT and scatterers location can be accurately estimated.
By contrast, when UT and scatterers are relatively far away from the BS as shown in Fig.~\ref{fig:location_far}, the accuracy of scatterers sensing is impaired and missing detection arises due to the non-uniform sampling grid pattern. 
{ From Table II, we can conclude that the scatterers localization errors have a significant impact on localization algorithms that rely solely on geometric information. Algorithms utilizing TDoA information, which is our proposed refined localization method, show more resilience to scatterers localization errors.}
{ Besides, it can be seen that in Fig.~\ref{fig:location_far}, there exist 6 MPCs but the estimated number of MPCs is 5.}  
However, its impact on UT location sensing accuracy is insignificant for the VAs-aided localization and refined localization schemes, benefiting from the adopted weighting strategy and high-precision TDoA measurement. 

Furthermore, Fig.~\ref{fig7:localization RMSE} numerically exhibits the UT sensing performance with the criteria $\text{RMSE}_{\theta}$ and $\text{RMSE}_{r}$.  
It can be observed that with the increase of pilot overhead, the UT location sensing accuracy witnesses sustained improvement until convergence, which indicates that superior channel estimation results contributes to better location sensing performance with a certain range.     
Owing to ultra-high angular resolution of UM-MIMO array, utilizing LoS information allows high-precision angle estimation, while its distance estimation performance is limited. This is due to the non-uniform spatial sampling lattice, where the grid interval is inversely proportional to the distance.  
Thus, the straightforward grid mapping with LoS information will inevitably result in worse distance resolution when the target distance increases. 
To remedy this, joint utilization of VAs can lead to distance estimation enhancement but at a slight cost of angle estimation performance degradation. 
Particularly worth mentioning is that the aforementioned proposed methods do not depend on the delay information and are thus also suitable for narrowband systems, which offers a new location sensing method for band-limited systems.

For the broadband systems, the available TDoA information of MPCs can further promote the localization accuracy to the sub-centimeter level.
Meanwhile, we investigate the location sensing performance in the absence of LoS link, where the number of NLoS MPCs is set to $L=6$. We can observe a slight loss in angle estimation accuracy and similar distance estimation performance, which validates the effectiveness of the proposed location sensing method under different channel conditions.
On this basis, we further plot the empirical cumulative distribution function (CDF) of $\text{RMSE}_{\theta}$ and $\text{RMSE}_r$ in Fig.~\ref{fig:localization RMSE CDF} with measurement length $P=72$.
Both of them have relatively large variances owing to the non-uniform spatial sampling grid. 
Take the proposed TDoA-assisted localization method for example, for the best case, $\text{RMSE}_{\theta}$ and $\text{RMSE}_r$ can reach $-50\, \rm dB$ and 
$-30\, \rm dB$, respectively.
While for the worst case, $\text{RMSE}_{\theta}$ and $\text{RMSE}_r$ degrade to 
$-23\, \rm dB$ and  $0\, \rm dB$.
Anyway, from the statistical average view, the satisfactory performance can still be achieved. 

{ Fig. \ref{fig:localization RMSE SNR} shows the localization performance comparison against different uplink SNR levels with the number of BS antennas and the number of measurements are set as $N_{\textrm{BS} }= 512$ and $P=80$, respectively. It can be seen from the figure that the performance of the localization get better as the uplink SNR increases. Besides, the results demonstrate that our proposed refined TDoA-assisted localization method consistently outperforms other approaches across different SNR levels, both in terms of angle and distance estimation, which demonstrates the superiority and the robustness of our methods under various noise conditions.}

{ Fig. ~\ref{fig:localization RMSE BS} shows the localization performance comparison against different numbers of BS antennas with the measurement length set as $P=80$. As evident from both subplots, all localization algorithms demonstrate gradually improving performance as the number of BS antennas increases. The improvement in localization accuracy with more antennas can be attributed to the enhanced spatial resolution and diversity provided by larger antenna arrays, which allow for more precise angle and distance estimation. Besides, methods utilizing TDoA information consistently outperform the method only utilizing the geometric information. The substantial performance improvement demonstrates the importance of incorporating TDoA information in localization problems}

\subsection{Beamforming Design Results}
   
\begin{figure}[t]	
	\centering	
	\includegraphics[width=0.5\columnwidth, keepaspectratio]{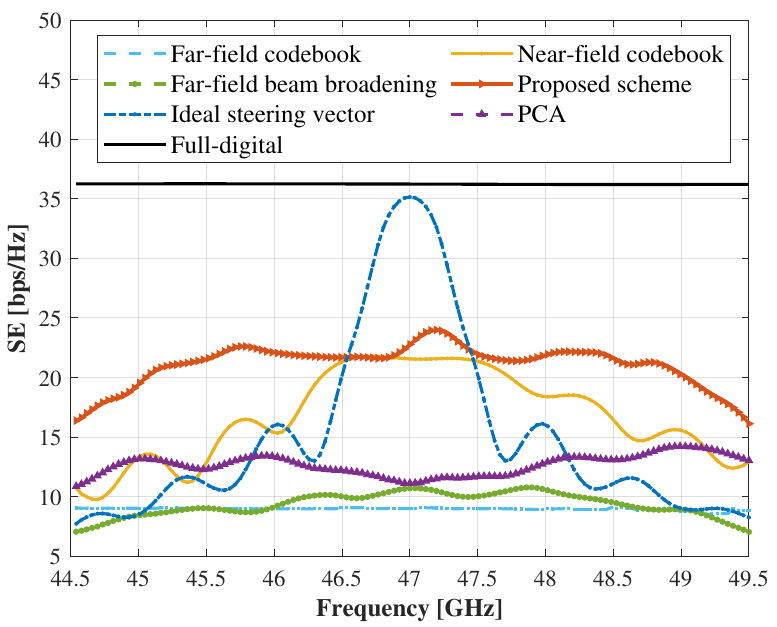}
	\captionsetup{font={color = {black}}, justification = raggedright,labelsep=period}
	\caption{SE comparison across the whole frequency band.}	
	\label{fig:precodiing frequency}
	\vspace{-5mm}	
\end{figure}     
   
\begin{figure}[t]
	\centering
	\subfigure[Case with LoS component]{
		\includegraphics[width=0.46\columnwidth, keepaspectratio]{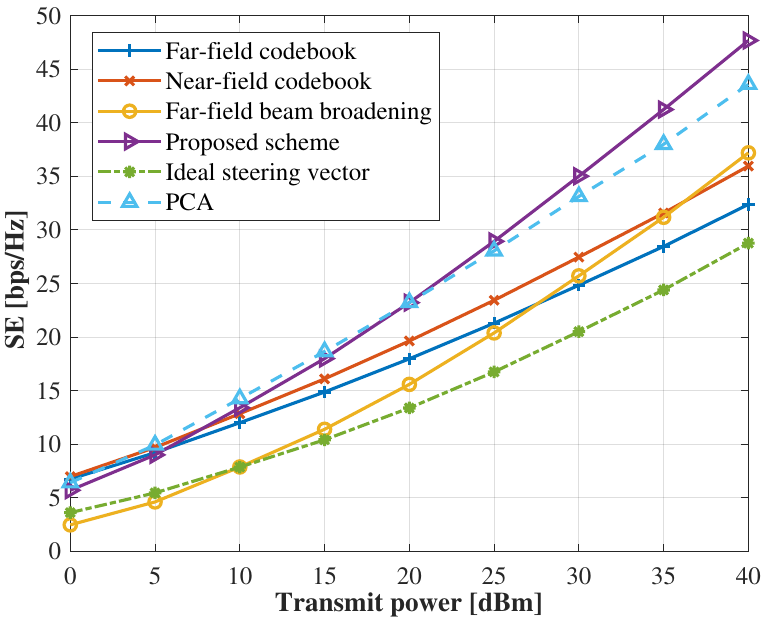}
	}	
	\subfigure[Case without LoS component]{		
		\includegraphics[width=0.46\columnwidth, keepaspectratio]{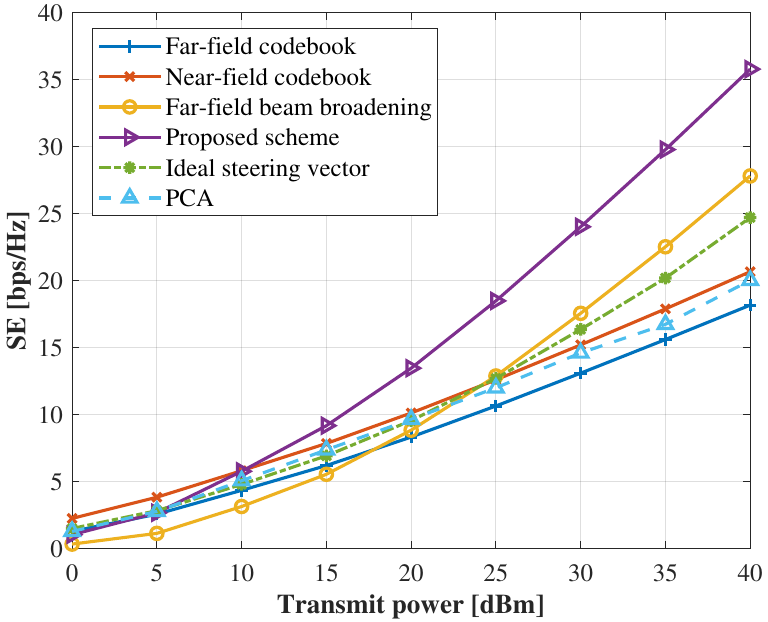}
	}
	\captionsetup{font={color = {black}}, justification = raggedright,labelsep=period}
	\caption{SE performance comparison with different transmit power $P_t^{\rm DL}$.}	
	\label{fig:precoding pow}	
	\vspace{-5mm}
\end{figure}   

\begin{figure}[t]	
	\centering
	\subfigure[Case with LoS component]{
		\includegraphics[width=0.46\columnwidth, keepaspectratio]{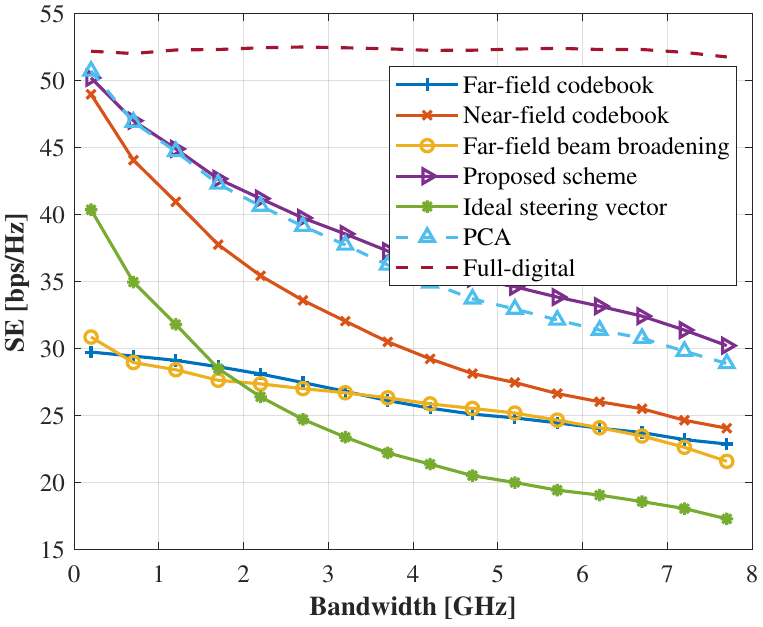}
	}	
	\subfigure[Case without LoS component]{		
		\includegraphics[width=0.46\columnwidth, keepaspectratio]{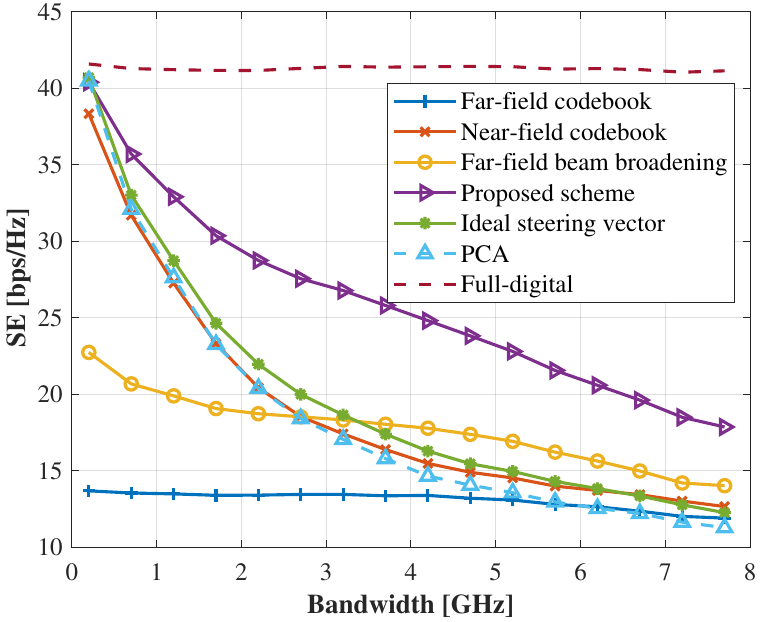}
	}
	\captionsetup{font={color = {black}}, justification = raggedright,labelsep=period}
	\caption{SE performance comparison with different bandwidth $\rm BW$.}	
	\label{fig:precoding bw}	
	\vspace{-5mm}
\end{figure}

Finally, we assess the effectiveness of beamforming design for data transmission by the SE criterion defined in \eqref{eq:SE}.
The baseline schemes for comparison include
\begin{enumerate}
	\item [(i)] DFT-codebook based spatially sparse precoding scheme \cite{omp-precoding}; 
	\item [(ii)] near-field codebook based beam focusing scheme \cite{Near-Field Beam Training}; 
	\item [(iii)] principal component analysis (PCA)-based beamforming scheme \cite{PCA}; 
	\item [(iv)] far-field beam boradening scheme \cite{gff}; 
	\item [(v)] ideal steering vector (for the center frequency) scheme. 
\end{enumerate}  
All of the schemes are based on the estimated channel and location sensing results, while the measurement length is set as $P=80$.

Fig.~\ref{fig:precodiing frequency} presents SE results calculated against different subcarriers, where the downlink transmit power is set to $P_t^{\rm DL} = 30 \, \rm dBm$. 
As we can see, the ideal steering vector at the center frequency utilized for analog phase shifts design can ensure approximately optimal beamforming design at the center frequency, but with severe loss at the marginal subcarriers.  
The traditional DFT codebook-based spatially sparse precoding scheme undergoes the worst performance across the entire frequency band due to its mismatch with the channel's HFBS feature. 
The near-field codebook beamforming scheme can remedy the aforementioned drawback but suffers from the problems of the beam-squint effect.
Similarly, the far-field beam-broadening scheme handles the probelms caused by the beam-squint effect partially but overlooks the coexisting hybrid-field effect.    
The proposed scheme jointly optimizes beamforming design by fully considering the HFBS effect and thus achieves the approximately consistent SE performance across the entire bandwidth.  

Fig.~\ref{fig:precoding pow} further compares SE performance with different transmit power in the statistical average view. In the presence of LoS component, the proposed method and the PCA scheme similarly achieve the best performance, while the proposed one maintains the superiority of lower computational complexity without matrix inversion for analog codebook design.
The proposed method further expands its advantages in the blocked LoS case and outperforms all other baseline schemes. 
In Fig.~\ref{fig:precoding bw}, we investigate the SE performance comparison with different bandwidth.  
When employing a relatively narrow band, all schemes except the far-field assumption based ones harvest satisfactory performance.   
With the increase of available bandwidth, the performance of all beamforming schemes  
deteriorates rapidly due to the practical hardware constraint of frequency-independent analog phase shifters.
Nevertheless, among them, the proposed one degrades most sluggishly and exhibits best robustness to the HFBS effect.

\section{Conclusion} 

In this paper, we not only overcomed but also exploited the HFBS nature of UM-MIMO systems and investigated a novel ILSC scheme. 
We commenced with proposing an ILSC signal frame structure consisting of uplink training  and data transmission stage.
In the uplink training stage, we utilized the designed projection matrix to achieve reliable sparse channel estimation under the HFBS effect, 
and further sensed the scatterers' location by extracting the angular and range parameters simultaneously from the spherical electromagnetic wave.
Taking the sensed scatterers as VAs, we conceived a w-LS method to coarsely estimate the 
the UT's location in line with its propagation geometry.
Leveraging the accurately estimated TDoA of MPCs, we further refined the location sensing results to a sub-decimeter level.
For the downlink data transmission beamforming design, we utilized the acquired channel and location information, and combined beam broadening and focusing strategy to 
mitigate the spectral efficiency degradation resulted from HFBS effect, and achieve constant beamforming gain as much as possible over the whole bandwidth and enhance SE.
Finally, we carried out simulation experiments to validate the superiority of the proposed ILSC scheme, and the results showed that the proposed schemes outperformed the baseline algorithms in terms of the channel estimation accuracy, location sensing performance, and communication SE.  

\end{spacing}

\begin{spacing}{0.90}

\end{spacing}

\end{document}